\documentclass[11pt]{article}

\makeatletter
\def\@biblabel#1{}
\makeatother

\renewcommand{\baselinestretch}{1.5}

\usepackage{fullpage}
\usepackage{amsfonts, subfigure}

\usepackage{caption} 
\usepackage{natbib}

\usepackage{graphicx}

\usepackage{mathrsfs}

\usepackage{amsmath}

\usepackage{amssymb}

\usepackage{float}

\usepackage[usenames]{color}
\usepackage[pdftex,hypertexnames=false,linktocpage=true]{hyperref}
\hypersetup{colorlinks=true,linkcolor=blue,anchorcolor=blue,citecolor=blue,filecolor=blue,urlcolor=blue,bookmarksnumbered=true,pdfview=FitB}

\def\bmu{\boldsymbol{\mu}}
\def\bSigma{\boldsymbol{\Sigma}}
\def\bomega{\boldsymbol{\omega}}

\def\bY{\mathbf{Y}}
\def\bX{\mathbf{X}}
\def\bZ{\mathbf{Z}}
\def\bS{\mathbf{S}}
\def\bD{\mathbf{D}}
\def\Sign{\mathbf{Sign}}
\def\Rank{\mathbf{Rank}}
\def\Mod{\mathrm{Mod}}

\begin{document}

\title{\bf Variable Importance Assessments and Backward Variable Selection for High-Dimensional Data}

\author{\setcounter{footnote}{0}
	Liuhua Peng\footnote{\baselineskip=11pt School of Mathematics and Statistics, University of Melbourne, Australia.   Email: liuhua.peng@unimelb.edu.au.},
	Long Qu, \setcounter{footnote}{2} Dan Nettleton\footnote{\baselineskip=11pt Department of Statistics \& Statistical Laboratory, Iowa State University, Ames, IA, USA.  Email: dnett@iastate.edu.}}

\date{}

\maketitle

\begin{abstract}
	Variable selection in high-dimensional scenarios is of great interested in statistics. One application involves identifying differentially expressed genes in genomic analysis. Existing methods for addressing this problem have some limits or disadvantages.  
	In this paper, we propose distance based variable importance measures to deal with these problems, which is inspired by the Multi-Response Permutation Procedure (MRPP). The proposed variable importance assessments can effectively measure the importance of an individual dimension by quantifying its influence on the differences between multivariate distributions. A backward selection algorithm is developed that can be used in high-dimensional variable selection to discover important variables. Both simulations and real data applications demonstrate that our proposed method enjoys good properties and has advantages over other methods.
\end{abstract}

\section{Introduction}

With the explosive and continued advancement of high-throughput biotechnologies, simultaneous measurement of more and more biological variables from any single experimental subject has become increasingly affordable and is frequently used in biomedical research. 
A distinguishing feature of these applications is that only a very limited number of experimental units (subjects) can be measured due to expense, leading to the ``small $n$, large $p$'' problem. Furthermore, the variables are expected to have very complex dependence structures governed by underlying biological processes that are not well characterized. In such a complex setting, only a small number of the most interesting and biologically meaningful variables or groups of variables are the primary targets for in-depth investigation.

High-dimensional variable selection is in great need in multiple scientific disciplines, particularly in modern genomics and personalized medicine. Microarray and RNA-seq technologies enable researchers to simultaneously measure thousands of potentially interesting variables. Identifying genes that differ in expression across two or more treatments or conditions is of great interest in genomic analysis. Identification of differentially expressed (DE) genes not only gives information about gene functionality, but also provides insight into the molecular genetic mechanisms underlying biological processes.

Although variable selection is not a new problem in statistics, existing statistical tools for variable selection in such high-dimensional contexts are still limited in capability. Current methods usually suffer from one or more of the following shortcomings: the number of selected variables cannot exceed the sample size; variable importance is evaluated based on a comparison of univariate marginal distributions; variables are selected in a forward manner; variable importance is based on distorted dependence structures that are not evident in the data; strong model assumptions (typically on the mean structure) are imposed during variable selection; and the selected variables can only capture certain aspects of dependency. Because of these limitations, there is a great gap between the needs of biological researchers and the capability of existing statistical tools for variable selection. This paper aims at avoiding or alleviating these aforementioned shortcomings.

The Multi-Response Permutation Procedure (MRPP) described by \citet{Mielke2007} is a powerful tool that can detect differences between multivariate distributions. The test statistic is based on a weighted average of within-treatment pairwise distances, and the testing procedure is carried out by permuting the observations. Moreover, under some mild conditions, the MRPP test is equivalent to the distance-based test proposed in \citet{SR2004}, which is inspired by the ``energy distance'' \citep{SR2013} and ``distance component analysis'' \citet{Rizzo2010}.  Distance-based methods such as the MRPP and energy statistics, have good features when dealing with multivariate and even high-dimensional problems, especially in capturing dependence structure among variables.

Inspired by former works, in this paper, we introduce distance-based variable importance measures for high-dimensional contexts that automatically take covariance structures into consideration. The importance measures are based on the idea of imposing a hypothetical perturbation on each dimension, and the importance is evaluated as the effect of the perturbation on the $p$-values of testing for differences among or between distributions. Furthermore, we propose a backward selection algorithm that can be used to select most important variables. By eliminating irrelevant dimensions iteratively, we can lower the dimensions of the data to alleviate the effect of high-dimensionality.  Examples in both real data and simulation studies show that our proposed method has good performance when detecting differentially expressed genes in genomic analysis.

The paper is organized as follows. Preliminaries are presented in Section \ref{sec:prelim}. Two importance measures are introduced in Section \ref{sec_MRPP_imp} and \ref{sec:imp_ed}. Section \ref{Sec_BS} introduces the proposed backward selection algorithm. Section \ref{sec:modified_MRPP} presents a modified MRPP. Section \ref{Sec_Real_data} gives an example of applying our method on a real data set. Simulation studies are shown in Section \ref{Sec_Simu}.

\section{Preliminaries}\label{sec:prelim}

\subsection{The Multi-Response Permutation Procedure (MRPP)}

The MRPP, a permutation method for testing equality of joint distributions, is described as a ``distance function approach'' by \citet{Mielke2007}. 
Consider a $K$-sample comparison experiment with an $R$-dimensional response vector. Let $\bY_i$ be the $i$-th observation of the $R$-dimensional response vector, with $r$-th element $Y_{i, r}$. The MRPP distance measure between observations $i$ and $j$ is usually chosen
to be Euclidean distance,
\begin{align}\label{eq:euclidean_norm}
	\Delta(i, j) = \sqrt{\sum_{r=1}^{R}\left(Y_{i,r}-Y_{j,r}\right)^2}=||\bY_i-\bY_j||.
\end{align}
Suppose we have a total of $N$ independent observations, each of which comes from exactly one of the $K$ treatments. Let $M_b(i)$ be the treatment label of observation $i$ under the $b$-th permutation of the $N$ observations, where $b$ is the factoradic number that indexes all $B = N!$ permutations of the observations, and $b = 0$ indicates the original assignment of treatment labels to observations. Further let $n_k$ be the sample size in the $k$-th treatment such that $N=\sum_{k=1}^K n_k$. The MRPP test statistic is
\begin{align}\label{eq:z_b_Delta}
	z_b(\Delta) = \sum_{k=1}^{K}C_k\left\{\frac{2}{n_k(n_k-1)}\sum_{(i,j)\in T_b(k)}\Delta(i,j)\right\},
\end{align}
where $C_k$ is the group weight usually chosen to be proportional to $n_k/N$ or $(n_k-1)/(N-K)$, and $T_b(k) = \{(i, j) : M_b(i) = M_b(j) = k, \ i < j, \ i = 1, \ldots, N, \ j = 1, \ldots, N\}$. The final permutation $p$-value for testing the null hypothesis of no distributional difference across the $K$ treatments is
\begin{align}\label{eq:MRPP_pvalue}
	p(\Delta) = \frac{1}{B}\sum_{b=0}^{B-1}I\left\{z_0(\Delta)\geq z_b(\Delta)\right\}.
\end{align}

Let $M$ be the sorted non-redundant set of sample sizes $\{n_i: i = 1, \ldots, K\}$ with the $j$-th element $m_j$, for $j = 1, \ldots, |M|$. Because all within treatment permutations are equivalent and exchanging treatment labels between treatment $i$ and $i'$ when $n_i = n_{i'}$ also results in the same test statistic, the support of the
$p$-value is the discrete set $\{b'/B': b' = 1, \ldots, B'\}$ where $B' = \frac{N!}{\prod_{i=1}^K n_i!\prod_{j=1}^{|M|} m_j!} $. When $B'$ is large, we may randomly sample $\tilde{B} \ll B'$ permutations from $B'$ non-equivalent permutations to save computing time. Whether using all $B'$ permutations or using a random subset of permutations, the type $I$ error rate is bounded above by $\alpha$ when the null hypothesis of distributional equality is rejected if and only if $p \leq \alpha$ \citep{Mielke2007}.

The MRPP has the advantage of recognizing, accounting for, and utilizing dependence information among the $R$-dimensions and capturing information about the joint distribution rather than only each marginal distribution. It has demonstrated good performance in the context of gene set testing \citep{Nettleton2008}. The associated multiple testing problem \citep{Liang2010} and variations of the test for more targeted hypotheses on variances \citep{Qu2010} have been addressed.

\subsection{Energy distance and distance components analysis}

Energy distance, proposed by \citet{SR2004}, is a measure of differences between two multivariate distributions. Suppose $\bX \sim F$ and $\bZ \sim G$ are two independent $R$-dimensional random vectors with finite means. The energy distance between $\bX$ and $\bZ$ is defined as
\begin{align*}
\mathcal{E}(\bX, \bZ) = 2E\|\bX-\bZ\|-E\|\bX-\bX'\|-E\|\bZ-\bZ'\|,
\end{align*}
where $\bX, \bX'\overset{\text{i.i.d.}}{\sim}F$ independent of $\bZ, \bZ'\overset{\text{i.i.d.}}{\sim}G$. A pleasant property of the energy distance is that $\mathcal{E}(\bX, \bZ)\geq0$ with equality to zero if and only if $\bX$ and $\bZ$ are identically distributed \citep{SR2013}. Suppose we have independent samples $\mathfrak{Y}_1=\{\bY_1, \ldots,\bY_{n_1}\}\sim F$ and $\mathfrak{Y}_2=\{\bY_{n_1+1}, \ldots,\bY_{n_1+n_2}\}\sim G$, the two-sample energy statistic corresponding to $\mathcal{E}(\bX, \bZ)$ is
\begin{align}\label{eq:energy_stat}
\mathcal{E}_{n_1,n_2}(\mathfrak{Y}_1, \mathfrak{Y}_2) = \frac{2}{n_1n_2}\sum_{i=1}^{n_1}\sum_{j=1}^{n_2}\|\bY_i-\bY_{n_1+j}\| - \frac{1}{n_1^2}\sum_{i=1}^{n_1}\sum_{j=1}^{n_1}\|\bY_i-\bY_{j}\| - \frac{1}{n_2^2}\sum_{i=1}^{n_2}\sum_{j=1}^{n_2}\|\bY_{n_1+i}-\bY_{n_1+j}\|.
\end{align}

The energy statistic $\mathcal{E}_{n_1,n_2}(\mathfrak{Y}_1, \mathfrak{Y}_2)$ has been used to test the equality of $F$ and $G$ \citep{SR2004}. The test can be implemented in a distribution free way by permuting the pooled sample $\{\mathfrak{Y}_1, \mathfrak{Y}_2\}$ to determine a reference distribution.

\citet{Rizzo2010} extended the two-sample energy statistic to multi-sample cases. The distance components (DISCO) analysis, viewed as a nonparametric analog of the classical analysis of variance (ANOVA), can be used to the multi-sample test of equal distributions. Consider a $K$-sample comparison experiment with $\mathfrak{Y}=\{\bY_1,\ldots,\bY_N\}$ as an observed independent sample where each observed vector comes from exactly one of the $K$ treatments. 
Following the notations in \citet{Rizzo2010}, the energy statistic between treatment $k$ and $k'$ is defined as
\begin{align*}
d_{\alpha}(\mathfrak{Y}_k, \mathfrak{Y}_{k'}) = \frac{n_kn_{k'}}{n_k+n_{k'}}\left[2g_{\alpha}(\mathfrak{Y}_k,\mathfrak{Y}_{k'})-g_{\alpha}(\mathfrak{Y}_k,\mathfrak{Y}_k)-g_{\alpha}(\mathfrak{Y}_{k'},\mathfrak{Y}_{k'})\right]
\end{align*}
where $\mathfrak{Y}_k$ is the sample of size $n_k$ from the $k$-th treatment and $$g_{\alpha}(\mathfrak{Y}_k,\mathfrak{Y}_{k'})=\frac{1}{n_kn_{k'}}\sum_{\{i:M_0(i)=k\}}\sum_{\{j:M_0(j)=k'\}}\|\bY_i-\bY_j\|^{\alpha},$$
with $\alpha\in(0,2)$ often chosen to be $1$. Then the between-sample and within-sample dispersions are defined as
\begin{align*}
\mathcal{S}_{\alpha}=\mathcal{S}_{\alpha}(\mathfrak{Y}_1, \ldots, \mathfrak{Y}_{K})=\sum_{1\leq k<k'\leq K}\left(\frac{n_k+n_{k'}}{N}\right)d_{\alpha}(\mathfrak{Y}_k, \mathfrak{Y}_{k'}),
\end{align*}
and
\begin{align*}
\mathcal{W}_{\alpha}=\mathcal{W}_{\alpha}(\mathfrak{Y}_1, \ldots, \mathfrak{Y}_{K})=\sum_{k=1}^{K}\frac{n_k}{2}g_{\alpha}(\mathfrak{Y}_k, \mathfrak{Y}_k),
\end{align*}
where both $\mathcal{S}_{\alpha}$ and $\mathcal{W}_{\alpha}$ are nonnegative and $\mathcal{S}_{\alpha}=0$ if and only if $\mathfrak{Y}_1=\cdots=\mathfrak{Y}_K$. The total dispersion is defined as
\begin{align*}
\mathcal{T}_{\alpha}=\mathcal{T}_{\alpha}(\mathfrak{Y}_1, \ldots, \mathfrak{Y}_{K})=\frac{N}{2}g_{\alpha}(\mathfrak{Y},\mathfrak{Y}),
\end{align*}
and we have the following DISCO decomposition for $K$-sample one-way design:
\begin{align*}
\mathcal{T}_{\alpha} = \mathcal{S}_{\alpha} + \mathcal{W}_{\alpha}.
\end{align*}
Furthermore, the DISCO $\mathcal{F}_{\alpha}$ ratio statistic for testing equal distributions is
\begin{align*}
\mathcal{F}_{\alpha}=\frac{\mathcal{S}_{\alpha}/(K-1)}{\mathcal{W}_{\alpha}/(N-K)},
\end{align*}
which is similar to the analysis of variance (ANOVA) $F$-statistic, but for testing distributional differences rather than mean differences. As for the two-sample test, the DISCO test can be implemented as a permutation test. In this paper, we drop the $\alpha$ subscript and only consider the case $\alpha=1$. 
That is, we use $\mathcal{S}$, $\mathcal{W}$, $\mathcal{T}$ and $\mathcal{F}$ to denote the corresponding between-sample and within-sample dispersion, the total dispersion and the DISCO $\mathcal{F}$ ratio statistic, respectively. See more discussions on the choice of $\alpha$ in \citet{Rizzo2010}.

\section{Importance measures based on MRPP}\label{sec_MRPP_imp}

The focus of the MRPP is on detecting differences between multivariate distributions. The procedure does not provide a measure of the importance of any one variable with respect to the information it contains about distributional differences. In this section, we introduce a method that ranks the importance of the $R$ variables and performs variable selection for the MRPP. Briefly, the ranking procedure consists of a hypothetical perturbation method that tilts each of the $R$ dimensions and a scheme for assessing the effects of such perturbations. Intuitively, the dimensions that lead to large differences in results under a small perturbation will be more influential and potentially more important than other dimensions.

Because the MRPP procedure relies on the distance measure $\Delta$, it is natural to consider a weighted Euclidean distance as an extension of the Euclidean distance for use in the MRPP. Let $\bomega$ be an $R$-vector with $r$-th element $\omega_r \geq 0$ being the weight for the $r$-th dimension. Then the weighted Euclidean distance between observations $i$ and $j$ is
\begin{align}\label{eq:weighted_euclidean_norm}
	\Delta_{\bomega}(i, j) = \sqrt{\sum_{r=1}^{R}\omega_r\left(Y_{i,r}-Y_{j,r}\right)^2}=||\bY_i-\bY_j||_{\bomega}.
\end{align}
When $\omega_r = 1$ for all $r = 1, \ldots, R$, \eqref{eq:weighted_euclidean_norm} is equivalent to \eqref{eq:euclidean_norm}. Note that introducing the weights is only conceptual and in practice $\omega_r$ can be always set to $1$. The advantage of using weights is that we can hypothetically increase or decrease some $\omega_r$ as a means of data perturbation for the purpose of evaluating variable importance. For example, setting $\omega_r = 0$ is equivalent to omitting dimension $r$ from analysis, which is similar to dropping a regressor in regression variable selection.

Given our method of perturbation, we now seek a measure of the effect of the perturbation. Because the end result for a permutation test is a permutation $p$-value, it is reasonable to consider how much the permutation $p$-value is changed by perturbation. However, because of the discreteness of the support of the permutation $p$-values defined in \eqref{eq:MRPP_pvalue}, if the perturbation in $\omega_r$ is too small, the permutation $p$-value $p(\Delta_{\bomega})$ may not change. On the other hand, we also want the perturbation to be as small as possible to faithfully reflect the original data set. To solve this conflict, we consider an approximation to the discrete permutation $p$-value in \eqref{eq:MRPP_pvalue} by a continuous $p$-value. Our choice is to treat the $B$ permutation test statistics as a random sample of size $B$ from an infinite population with a cumulative distribution function (CDF) $F$ and density $f$, and to apply the kernel method to estimate $F$ and $f$. Using the Gaussian kernel with bandwidth $h$, the kernel estimate of $F$ is
\begin{align*}
	\hat{F}(z) = \frac{1}{B}\sum_{b=0}^{B-1}\Phi\left\{\frac{z-z_b(\Delta_{\bomega})}{h}\right\},
\end{align*}
where $\Phi$ is the CDF for the standard normal distribution. The continuous approximation to the discrete $p$-value in (\ref{eq:MRPP_pvalue}) is then given by $ \tilde{p}(\Delta_{\bomega}) = \hat{F}\{z_0(\Delta_{\bomega})\} $, evaluated at $\omega_r = 1$ for all $r = 1, \ldots, R$. The choice of bandwidth $h$ is well known to be crucial for the performance of kernel density estimation \citep{Scott1992,Wand1995}. We defer the discussion of its choice to next subsection.

We can now compute the importance, $\iota_r$, of variable $r$ as the partial derivative of the continuous $p$-value $\tilde{p}(\Delta_{\bomega})$ with respect to weight $\omega_r$, evaluated at $\omega_r = 1$ for $r = 1, \ldots, R$. Specifically, the importance (or the influence) of the $r$-th dimension is computed as
\begin{align}
	\iota_r = \left. \frac{\partial \tilde{p}(\Delta_{\bomega})}{\partial \omega_r}\right |_{\bomega=\textbf{1}} = \frac{1}{Bh}\sum_{b=0}^{B-1}\phi\left\{\frac{z_0(\Delta)-z_b(\Delta)}{h}\right\}\left\{z_0(\nabla_r)-z_b(\nabla_r)\right\},
\end{align}
where $\phi$ is the standard normal density and
$$ \nabla_r(i,j) = \left. \frac{\partial \Delta_{\bomega}(i,j)}{\partial \omega_r}\right |_{{\bomega}=\textbf{1}} = \frac{(Y_{i,r}-Y_{j,r})^2}{2\Delta(i,j)}$$
for all $r = 1, \ldots, R$. If the derivate is negative, then increasing the weight will decrease the $p$-value, i.e., the $r$-th dimension is important. On the other hand, if the derivative is positive, then increasing the weight will increase the $p$-value, and focusing more on the $r$-th dimension diminishes the significance of the MRPP.


Other than taking account of the dependency and robustness to normality, another important advantage of MRPP -- allowing the dimensionality to exceed the sample size -- is also inherited by our variable ranking procedure. This ensures that our proposed importance measure is applicable for the high-dimensional context. Moreover, when we measure the importance of the $r$-th dimension, the remaining $R-1$ dimensions have not been excluded from the data, even if $R-1>N$. This allows backward variable selection procedures to be possible in high dimensions and is particularly advantageous compared to marginal screening procedures, for example, the marginal Pearson correlation screening \citep{Fan2008}, the marginal distance correlation screening \citep{LiZhongZhu2012}, the marginal maximal information coefficient \citep{Reshef2011,Speed2011,Gorfine2012,Simon2012}, or the marginal empirical likelihood screening \citep{Chang2013, Chang2016}.

Compared to similar permutation methods that permute each dimension separately to assess variable importance, e.g., as in the random forest procedure \citep{Breiman2001}, our method does not distort the inter-relationship between the variable under consideration and the remaining variables. Thus, our method is more faithful to the observed data and reflects the true importance of a variable in the joint distribution of all response variables, rather than in the distribution where the variable under consideration and the remaining variables are artificially decorrelated through permutation. This is very important in terms of biological interpretations. In molecular biology, it is well known that intracellular environment is crucial and genes interact with each other in a complex manner. The same gene may have different functions depending on how related genes are expressed. Therefore, a statistical procedure that assesses whether a gene is important must account for the expression levels of other genes. Because our method does this accounting, it provides a potentially more meaningful solution to molecular biology researchers.

\subsection{Importance measures under small discrete perturbations}

While $\iota_r$ measures variable importance under hypothetical infinitesimal perturbation, it can be approximated by actual discrete analogs. With $\bomega_{(-r)}=\textbf{1}$, the partial derivative of $\tilde{p}(\Delta_{\bomega})$ at $w_r=1$ can be approximated by the slope of nearby secant lines with $w_r=0$ or $w_r=2$. This gives the following backward, forward, and central finite difference approximants of $\iota_r$:
\begin{align*}
	\iota_r & \approx \tilde{p}(\Delta) - \tilde{p}(\Delta^{(-r)}) \triangleq \iota_r^{-} \\
			& \approx \tilde{p}(\Delta^{(+r)}) - \tilde{p}(\Delta) \triangleq \iota_r^{+} \\
			& \approx \frac{1}{2}\big\{\tilde{p}(\Delta^{(+r)}) - \tilde{p}(\Delta^{(-r)})\big\} \triangleq \iota_r^{\pm},
\end{align*}
where
\begin{align*}
	\Delta^{(-r)}(i, j) = \sqrt{\sum_{s=1,s\neq r}^{R}\left(Y_{i,s}-Y_{j,s}\right)^2}
\end{align*}
and
\begin{align*}
	\Delta^{(+r)}(i, j) = \sqrt{\sum_{s=1}^{R}\left(Y_{i,s}-Y_{j,s}\right)^2+(Y_{i,r}-Y_{j,r})^2}
\end{align*}
are, respectively, the Euclidean distance computed without the $r$-th dimension or with an extra $(R+1)$-th dimension that is identical to dimension $r$. We can call them as the drop-1-variable and add-1-variable methods.

These approximations can be used as variable importance measures with intuitive interpretations similar to $\iota_r$. If the $r$-th variable is important, dropping it tends to produce larger $p$-values and double weighting it tends to produce smaller $p$-values. An advantage of such measures is the avoidance of choosing a bandwidth
$h$. However, the permutation $p$-values are inherently discrete. Variable ranking using these discrete measures might produce many ties.


Note that, in regression variable selection problems, the drop-1-variable and the keep-1-variable procedures are common. But the add-1-variable approach is rare, partly because complete collinearity introduced by the added variable is often considered an anomaly in regression. But distance based methods do not suffer from this issue.

\subsection{Choice of $h$ in $\iota_r$}

Since $\iota_r$ depends on the kernel smoothing of permutation statistics, existing methods for choosing $h$ developed under kernel smoothing contexts might be applied. For example, we may select an asymptotically optimal $h$ to minimize the mean squared error of $\hat{f}=\hat{F}'$ or of $\hat{F}$ at the fixed point $z_b(\Delta)$ \citep{Scott1992,Wand1995}. But such optimality is unjustified because these optimal results were developed under the assumption of an infinite population, whereas the permutation statistics form a finite discrete set.

Another heuristic choice of $h$ is to maximize $\sum_{b=0}^{B}\{\hat{f}'(z_b(\Delta))\}^2$. The rational is that if $h$ is too small, then $\hat{f}$ will approximately be a set of non-overlapping spikes located each of $z_b(\Delta)$, with $\hat{f}'(z_b(\Delta))\approx0$ for all $b$. If $h$ is too large, then $\hat{f}$ will be a very wide unimodal bell-curve that is nearly flat over the finite range of $z_b(\Delta)$ statistics, and $\hat{f}'(z_b(\Delta))$ is still close to $0$ for all $b$. Thereby, we might seek an intermediate $h$ that avoids such extreme choices by letting most $\hat{f}'(z_b(\Delta))$ be sufficiently different than $0$.

The above choices of $h$ all suffer from \textit{ad hoc} subjectivity to some degree. They are implemented in the \textbf{R} package \textbf{MRPP} for users to explore. However, we prefer the following more objective methods. Because the goal of choosing $h$ is to compute a derivative $\iota_r$, a good $h$ should give a good $\iota_r$ that is close to its data-dependent approximations $\iota_r^{-}$, $\iota_r^{+}$ and $\iota_r^{\pm}$. Thus, we may choose $h$ to minimize any of the following sum of squared errors,
\begin{align*}
	\sum_{r=1}^{R}(\iota_r-\iota_r^{-})^2, & ~~\sum_{r=1}^{R}(\iota_r-\iota_r^{+})^2, \\
	\sum_{r=1}^{R}(\iota_r-\iota_r^{\pm})^2, & ~~\sum_{r=1}^{R}(\iota_r-\iota_r^{-})^2+\sum_{r=1}^{R}(\iota_r-\iota_r^{+})^2.
\end{align*}
Because the latter two choices seek good approximation to both the drop-1-variable
and add-1-variable method, they are our preferred methods.


\section{Importance measures based on the energy distance}\label{sec:imp_ed}

The variable importance measure $\iota_r$ 
is derived by using a kernel-smoothing approximation to the permutation distribution of the MRPP test statistic. As argued in 
Section \ref{sec_MRPP_imp},
we believe $\iota_r$ is an intuitively appealing measure of importance that avoids drawbacks of other approaches. 
However, $\iota_r$ has it own drawbacks. First, the reliance on the permutation distribution of the MRPP test statistic involves nontrivial computational expense in high-dimensional problems. Second, the need to specify a bandwidth parameter $h$ is an inconvenience. Third, although the rationale of developing $\iota_r$ is appealing, the population analog that $\iota_r$ tries to approximate is not immediately clear, impeding the study of its theoretical properties. In this 
section, we propose an alternative variable importance measure that takes $\iota_r$ as a starting point and attempts to eliminate its drawbacks while maintaining its appealing features. 

\subsection{$h\to\infty$ in $\iota_r$}

The first two drawbacks of $\iota_r$ can actually be avoided by a special choice of $h$. Note that, as a variable relative importance measure, any common factor outside of the summation in the $\iota_r$ equation does not affect variable ranking. That is, $\{\iota_r\}_{r=1}^R$ and $\{\iota_rh/\phi(0)\}_{r=1}^R$ rank variables identically. Letting $h\to\infty$, we have
\begin{align*}
	\frac{\iota_rh}{\phi(0)}  \to \frac{1}{B}\sum_{b=0}^{B-1}\{z_0(\nabla_r)-z_b(\nabla_r)\}  = z_0(\nabla_r)-\frac{1}{B}\sum_{b=0}^{B-1}z_b(\nabla_r) \triangleq \tau_r,
\end{align*}
where $\tau_r$ obviously avoids the otherwise inconvenient choice of $h$.

By inspecting the above equation, we see that $\tau_r$ is a centered MRPP statistic using $\nabla_r$ as the distance measure, centered by its permutation mean. For $\nabla_r$ itself, the numerator specifically measures the contribution of the $r$-th dimension to the within-treatment squared distance, and the denominator re-weights this contribution relative to the overall within-treatment distance across all dimensions. Hence, if the $r$-th variable is important and if we use $\nabla_r$ as a distance measure, $z_0(\nabla_r)$ will tend to be more significant compared to its permutation distribution. The centering performed by $\tau_r$ can be thought as a means to make the permutation distribution more comparable across different variables. An more computationally intensive alternative that fully achieves
comparability is $p(\nabla_r)$, although the latter suffers more from discreteness than $\tau_r$.

Furthermore, standard combinatorics arguments show that
\begin{align}\label{eq:doule_sum}
	\frac{1}{B}\sum_{b=0}^{B-1}z_b(\nabla_r) = \frac{2}{N(N-1)}\sum_{1\leq i<j\leq N}\nabla_r(i,j),
\end{align}
which is a result analogous to the permutation moments of MRPP statistics \citep{Mielke2007}. This is remarkable because it is free of group weights $C_k$, and it drastically decreases the computational complexity of $\tau_r$ from order $B\times N^2$ to order $N^2$, i.e., the computationally expensive permutations can be completely avoided irrespective to how $C_k$ is chosen.

Treating $\nabla_r$ as the distance, the double summation in \eqref{eq:doule_sum} also enjoys the interpretation as an MRPP statistic computed under the null hypothesis that all observations are i.i.d., i.e., all $K$ groups can be pooled as a single homogeneous group. Therefore, $\tau_r$ can also be interpreted as the difference between the MRPP statistics under the null and the alternative hypotheses, when the effect of dimension $r$ is concentrated using $\nabla_r$ as the distance measure.

\subsection{Relation to energy distance and distance component analysis}

So far, we have focused on the permutation test context that treats observations as fixed quantities, or a context that conditions on a minimal sufficient statistic of the random data. We now consider the unconditional situation where data are treated as random variables.

For ease of exposition, we temporarily assume that $K = 2$. Suppose we choose $C_k = n_k/N$ for $k = 1, 2$. It follows that
\begin{align*}
	\tau_r & = -\bigg\{\frac{2}{N(N-1)}\sum_{1\leq i<j\leq N}\nabla_r(i,j)-\sum_{k=1}^{2}\frac{2}{N(n_k-1)}\sum_{(i,j)\in T_0(k)}\nabla_r(i,j)\bigg\} \\
			 & = -\frac{n_1n_2}{N(N-1)}\bigg\{\frac{2}{n_1n_2}\sum_{(i,j):M_0(i)\neq M_0(j), i<j}\nabla_r(i,j)-\sum_{k=1}^{2}\frac{2}{n_k(n_k-1)}\sum_{(i,j)\in T_0(k)}\nabla_r(i,j)\bigg\}.
\end{align*}
Because the MRPP statistic is based on $U$-statistics, taking the unconditional expectation of $\tau_r$ gives
\begin{align*}
	E(\tau_r) = -\frac{n_1n_2}{N(N-1)}\left[2E\{\nabla_r(i_1,i_2)\}-E\{\nabla_r(i_1,i_1')\}-E\{\nabla_r(i_2,i_2')\}\right],
\end{align*}
where $i_1$ and $i_1'$ are two different indices of observations from the first treatment, and $i_2$ and $i_2'$ are two different indices of observations from the second treatment.

To see what $E(\tau_r)$ evaluates to, consider the weighted energy distance $\mathcal{E}_{\bomega}$, with Euclidean distance $\Delta$ replaced by weighted Euclidean distance $\Delta_{\bomega}$ in the definition of energy distance. Following previous arguments, a population variable importance measure based on the energy distance can be chosen as the partial derivative of $\mathcal{E}_{\bomega}$ with respect to $\omega_r$, evaluated at $\bomega=\mathbf{1}$. Assuming the exchangeability of differentiation and integration, we have
\begin{align*}
	\epsilon_r \triangleq \frac{\partial\mathcal{E}_{\bomega}(\bY_{i_1},\bY_{i_2})}{\partial\omega_r}\bigg|_{\bomega=\mathbf{1}} & = 2E\{\nabla_r(i_1,i_2)\}-E\{\nabla_r(i_1,i_1')\}-E\{\nabla_r(i_2,i_2')\} \\
	& = -\frac{N(N-1)}{n_1n_2}E(\tau_r).
\end{align*}

In other words, rescaled $N(1-N)\tau_r/(n_1n_2)$ is an unbiased estimator of population variable importance $\epsilon_r$. As the rescaling factor is free of $r$, scaling does not affect variable ranking. This provides theoretical justification for using $\tau_r$ as the variable importance measure. If the $r$-th dimension is important, we should expect an increase in energy distance between treatments when the weight $\omega_r$ is increased by an infinitesimal amount and vice versa.

For a general $K$-sample problem, the importance measure $\tau_r$ is closely related with the between-sample dispersion $\mathcal{S}_{\alpha}$ in the DISCO analysis. Define
\begin{align*}
	\epsilon_r(\bY_{i_k}, \bY_{i_{\ell}}) = 2E\{\nabla_r(i_k,i_{\ell})\}-E\{\nabla_r(i_k,i_k')\}-E\{\nabla_r(i_{\ell},i_{\ell}')\},
\end{align*}
for $r=1,\ldots,R$, $1\leq k, \ell\leq K$. Here $i_k$ and $i_k'$ are two different indices of observations from the $k$-th treatment, and $i_{\ell}$ and $i_{\ell}'$ are two different indices of observations from the $\ell$-th treatment. Then by choosing $C_k=n_k/N$ for $k=1,\ldots,K$,
\begin{align*}
	E(\tau_r) = & \sum_{k=1}^{K}\frac{n_k}{N}E\{\nabla_r(i_k,i_k')\}-\frac{1}{N(N-1)}\left[\sum_{k=1}^{K}n_k(n_k-1)E\{\nabla_r(i_k,i_k')\}+\sum_{k\neq \ell}n_kn_{\ell}E\{\nabla_r(i_k,i_{\ell})\}\right] \\
	= & -\frac{1}{N(N-1)}\sum_{1\leq k<\ell\leq K}n_kn_{\ell}\epsilon_r(\bY_{i_k}, \bY_{i_{\ell}}).
\end{align*}
On the other hand, the population version of $\mathcal{S}$ is
\begin{align*}
	\mathfrak{S}(\bY_{i_1},\ldots,\bY_{i_K}) = \frac{1}{N}\sum_{1\leq k<\ell\leq K}n_kn_{\ell}\mathcal{E}(\bY_{i_k}, \bY_{i_{\ell}}).
\end{align*}
Now replacing the energy distance $\mathcal{E}$ by its weighted version $\mathcal{E}_{\bomega}$ in $\mathfrak{S}(\bY_{i_1},\ldots,\bY_{i_K})$, we have
\begin{align*}
	E(\tau_r) = -(N-1)\frac{\partial\mathfrak{S}_{\bomega}(\bY_{i_1},\ldots,\bY_{i_K})}{\partial\omega_r}\bigg|_{\bomega=\mathbf{1}}.
\end{align*}
So $\tau_r$ is $U$-statistic based with expectation proportional to the partial derivative of $\mathfrak{S}_{\bomega}(\bY_{i_1},\ldots,\bY_{i_K})$ with respect to $\omega_r$, evaluated at $\bomega=\mathbf{1}$, which indicates the importance of dimension $r$.

Actually, we can use a $U$-statistic version of $\mathcal{S}$ such that $\mathcal{S}$ is an unbiased estimator of $\mathfrak{S}(\bY_{i_1},\ldots,\bY_{i_K})$, and this can be achieved by redefining
$$g_{U}(\mathfrak{Y}_k,\mathfrak{Y}_{k})=\frac{2}{n_k(n_k-1)}\sum_{(i,j)\in T_0(k)}\|\bY_i-\bY_j\|,$$
and replacing $g_{\alpha}(\mathfrak{Y}_k,\mathfrak{Y}_{k})$ with $g_{U}(\mathfrak{Y}_k,\mathfrak{Y}_{k})$ in $\mathcal{S}$. Define
\begin{align*}
	\mathcal{S}_{U}=\mathcal{S}_{U}(\mathfrak{Y}_1, \ldots, \mathfrak{Y}_{K})=\sum_{1\leq k<k'\leq K}\left(\frac{n_k+n_{k'}}{N}\right)d_{U}(\mathfrak{Y}_k, \mathfrak{Y}_{k'}),
\end{align*}
where $d_{U}(\mathfrak{Y}_k, \mathfrak{Y}_{k'}) = \frac{n_kn_{k'}}{n_k+n_{k'}}\left[2g_{\alpha}(\mathfrak{Y}_k,\mathfrak{Y}_{k'})-g_{U}(\mathfrak{Y}_k,\mathfrak{Y}_k)-g_{U}(\mathfrak{Y}_{k'},\mathfrak{Y}_{k'})\right]$ for $k\neq k'$. If we replace $\|\bY_i-\bY_j\|$ with $\|\bY_i-\bY_j\|_{\bomega}$ in $\mathcal{S}_U$, simple algebra shows that
\begin{align*}
	\tau_r = -(N-1)\frac{\partial\mathcal{S}_{U, \bomega}(\mathfrak{Y}_1, \ldots, \mathfrak{Y}_{K})}{\partial\omega_r}\bigg|_{\bomega=\mathbf{1}},
\end{align*}
which reveals the close relationship between the importance measure $\tau_r$ and the DISCO analysis.

Since $\tau_r$ and $\iota_r$ only differ on the choice of $h$, i.e., the usual average compared to a weighted average, we expect that $\iota_r$ also has similar properties in terms of ranking variables and is approximately proportional to $\epsilon_r$ on average. Indeed, from empirical studies not detailed here, $\iota_r$ and $\tau_r$ tend to be highly correlated.

Practically, if variable ranking is performed after an initial MRPP test, 
we slightly prefer using $\iota_r$ as the variable importance measure, for its closer agreement with the initial MRPP result. On the other hand, if variable selection is performed during a permutation test, 
we prefer using $\tau_r$, for its computational efficiency.
Because $\tau_r$ is computationally less expensive and does not require specification of a bandwidth parameter $h$, we use $\tau_r$ as our measure of variable importance throughout the subsequent sections of this paper.

\section{Backward Selection} \label{Sec_BS}

Our variable importance measure quantifies how one variable plays a role in the difference between two distributions. When dealing with high-dimensional distributions, the importance of individual dimensions can be obscured by irrelevant dimensions whose joint distribution is identical across treatment groups.  To eliminate such dimensions and focus attention on the most important variables, we propose a backward variable selection algorithm that can trim away irrelevant variables in a stepwise manner.

Our backward selection algorithm is defined as follows.  Let $\bS(\ell)$ be the indices of the selected variables at iteration $\ell$.  Let $\bD(\ell)$ be the indices of the deleted variables at iteration $\ell$.   Initialize $\bS(0)=\{1, \ldots, R\}$ and $\bD(0) = \emptyset$.  For $\ell \geq 1$, perform the following steps:

\begin{enumerate}
	\item For each $r\in\bS(\ell-1)$, let $\tau_{\ell,r}$ be the measure of variable importance for dimension $r$ when data vectors consist only of variables indexed by $\bS(\ell -1 )$.  Set $\Gamma(\ell) = \{\tau_{\ell,r}: r \in \bS(\ell-1)\}$.

	\item Let $s_{\ell, r} = 1 \times{\rm sign}(\tau_{\ell,r})$ for all $r \in \bS(\ell-1)$ and $s_{\ell, r} = 1$ for all $r \in \bD(\ell -1)$.  Let $\Sign(\ell) = \{s_{\ell,r}: r=1,\ldots, R\}$.

	\item For all $r \in \bS(\ell-1)$, let $\gamma_{\ell, r}$ be the rank of $\tau_{\ell,r}$ in $\Gamma(\ell)$.  For all $r \in \bD(\ell -1)$, let $\gamma_{\ell, r} = R-\tilde{\ell}+1$, where $\tilde{\ell}$ is the number of the iteration when the $r$-th variable was moved from the selected set to the deleted set (see step 4 below).  Let $\Rank(\ell) = \{\gamma_{\ell,r}: r=1,\ldots, R\}$.
 
	\item Find $\max \Gamma(\ell)$, and let $d(\ell)$ be the index corresponding to the maximum element of $\Gamma(\ell)$.

	\begin{enumerate}
		\item If $\max \Gamma(\ell) \geq 0$, compute the $p$-value of the MRPP test of distributional equality between treatment groups based only on the variables whose indices are in the set $\bD(\ell-1)\cup \{d(\ell)\}$.  If the $p$-value is less than a user-chosen threshold for significance, set $\bS(\ell)=\bS(\ell-1)$, $\bD(\ell)=\bD(\ell -1)$, and $L=\ell$ and stop iterating. Otherwise, set $\bS(\ell)=\bS(\ell-1)\setminus d(\ell)$ and $\bD(\ell)=\bD(\ell -1) \bigcup \{d(\ell)\}$ and continue iterating. 

		\item If $\max \Gamma(\ell) < 0$, set $\bS(\ell)=\bS(\ell-1)$, $\bD(\ell)=\bD(\ell -1)$, and $L=\ell$ and stop iterating.
	\end{enumerate}

\end{enumerate}

By using the results obtained from the backward selection, there are several approaches that can be used to make decisions about importance of variables. The first issue is to assess which variables are important. We can deal with this problem in two ways.  The first intuitive approach is to declare all variables with indices in the set $\bS(L)$ to be important and all variables indexed by $\bD(L)$ to be unimportant. A second method is based on the signs recorded in $\Sign(\ell)$ for $\ell=1,\ldots,L$. Ideally, each important variable will have a negative sign for each iteration, but random variation in the importance measures can lead to positive signs in some iterations for some important variables. Thus, it may make sense to consider variable $r$ important if $s_{\ell,r}<0$ for some large proportion (for example, 80\%) of iterations $\ell=1,\ldots, L$. We denote this set of important variables determined by the sign vectors as $\bS_{\Sign}(L;\delta)$, where $\delta$ is a threshold that specifies the percentage of negative signs needed for one variable to be classified as important. Furthermore, by calculating the average rank of variables according to the ranks contained in $\{\Rank(\ell), \ell = 1, \cdots, L\}$, we can compare the relative importance of variables based on their average ranks.


One motivation of doing backward selection is that, by iteratively deleting variables that are not important, we can reduce the dimensions of the data to relieve the effect of ``The Curse of Dimensionality''. This method tends to work well, especially when the covariance structures of the data vectors are complicated. Examples that illustrate this point will be given in Section \ref{Sec_Simu}.

\section{A Modified MRPP}\label{sec:modified_MRPP}

The backward selection algorithm proposed in the previous section can be used as a follow-up procedure to identify important variables when the original MRPP test detects a difference in multivariate distributions among treatment groups. In this section, we explain how our backward selection procedure can alternatively be used prior to the original MRPP test to concentrate attention of the most important subset of the variables that contains information about potentially lower-dimensional multivariate distribution differences embedded within high-dimensional data vectors. The procedure is defined as follows.
\begin{enumerate}
	\item Starting with the original dataset, perform backward selection to obtain the $R_0$ variables judged to be most important. $R_0$ can be determined by the cardinality of $\bS(L)$ or $\bS_{\Sign}(L,\delta)$ described in Section \ref{Sec_BS}. Alternatively, $R_0$ can be pre-selected. Compute the MRPP test statistic given in \eqref{eq:z_b_Delta} using only the $R_0$ variables chosen by backward selection.  Use $z_{0, bs}(\Delta_{R_0})$ to represent the value of the test statistic.
	
	\item  For the $b$-th permutation of the original dataset, where $b$ is the factoradic number that indexes all $B = N!$ permutations of the observations, do backward selection on the permuted data to select $R_0$ variables. After backward selection, calculate the MRPP test statistic only with the $R_0$ variables selected from the permuted data. Use $z_{b, bs}(\Delta_{R_0})$  to represent the value of the test statistic for permutation $b$.
	
	\item  The modified MRPP $p$-value is defined as $p_{bs}(\Delta) = \frac{1}{B}\sum_{b=0}^{B-1}I\left\{z_{0, bs}(\Delta_{R_0})\geq z_{b, bs}(\Delta_{R_0})\right\}$.
\end{enumerate}

This testing procedure is similar to the original MRPP, but instead of using all $R$ dimensions, we impose backward selection to focus on the variables that carry the strongest signal for distributional differences. For each permuted data, we select the same number of variables ($R_0$) as were selected for the original data so that the $p$-value $p_{bs}(\Delta)$ is derived by comparing the average of within-group pairwise distances based on data vectors of constant dimensionality. Moreover, because the permutation $p$-value involves comparing MRPP test statistics computed from varying subsets of the original $R$ variables, it is important to standardize each variable prior to conducting this modified MRPP test.

It has been discovered that with fixed number of signal-bearing dimensions, the power of the original MRPP and the test based on energy distance \citep{SR2004} decreases with increasing dimension \citep{Ramdas2015}. The modified MRPP can alleviate the power drop by focusing on the subset of important variables. Simulation studies presented in Section \ref{Sec_Simu} show that the size of modified MRPP can be well controlled and that the power of the modified procedure can exceed that of the original MRPP test on all $R$ dimensions, especially when there are a relatively small number of variables responsible for multivariate distributional differences.

\section{Real Data Analysis}\label{Sec_Real_data}

The MRPP has been used to detect differentially expressed gene sets in the analysis of microarray gene expression data in \citet{Nettleton2008}. To demonstrate the usefulness of our proposed variable selection method, we performed backward selection based on our variable importance measure on a real microarray data set and compared it with other methods.

We use the \textbf{ALL} dataset which consists of transcript abundance measurements on $12625$ genes for $128$ different individuals with acute lymphoblastic leukemia (\textbf{ALL}). As described in \citet{Hahne2008}, two subsets of interest in the data are $40$ individuals with B-cell tumors that carry the BCR/ABL mutation and $35$ individuals with B-cell tumors that have no observed cytogenetic abnormalities.  We restrict our attention to an analysis of these 75 samples.  As suggested in \citet{Hahne2008}, we consider a subset of 2149 genes that shows the greatest variation in transcript abundance levels across the 75 samples.  Among these 2149 genes is a set of $196$ genes associated with the gene ontology (GO) term ``positive regulation of transcription from RNA polymerase II promoter'' that we use as an example to illustrate our approaches.

For ease of reference, we will refer to the 40 samples with the BCR/ABL mutation as group 1 and the 35 without cytogenic abnormalities as group 2. The $p$-value of the MRPP test for a difference between between groups 1 and 2 with respect to the $196$-dimensional multivariate gene expression distribution is $0.002$ based on 1000 permutations.  This small $p$-value provides significant evidence of a distributional difference between groups 1 and 2 but provides no information about which of the $196$ genes may be primarily responsible for the difference.   

\begin{figure}[h!]
	\centering
	\includegraphics[width=0.8\linewidth]{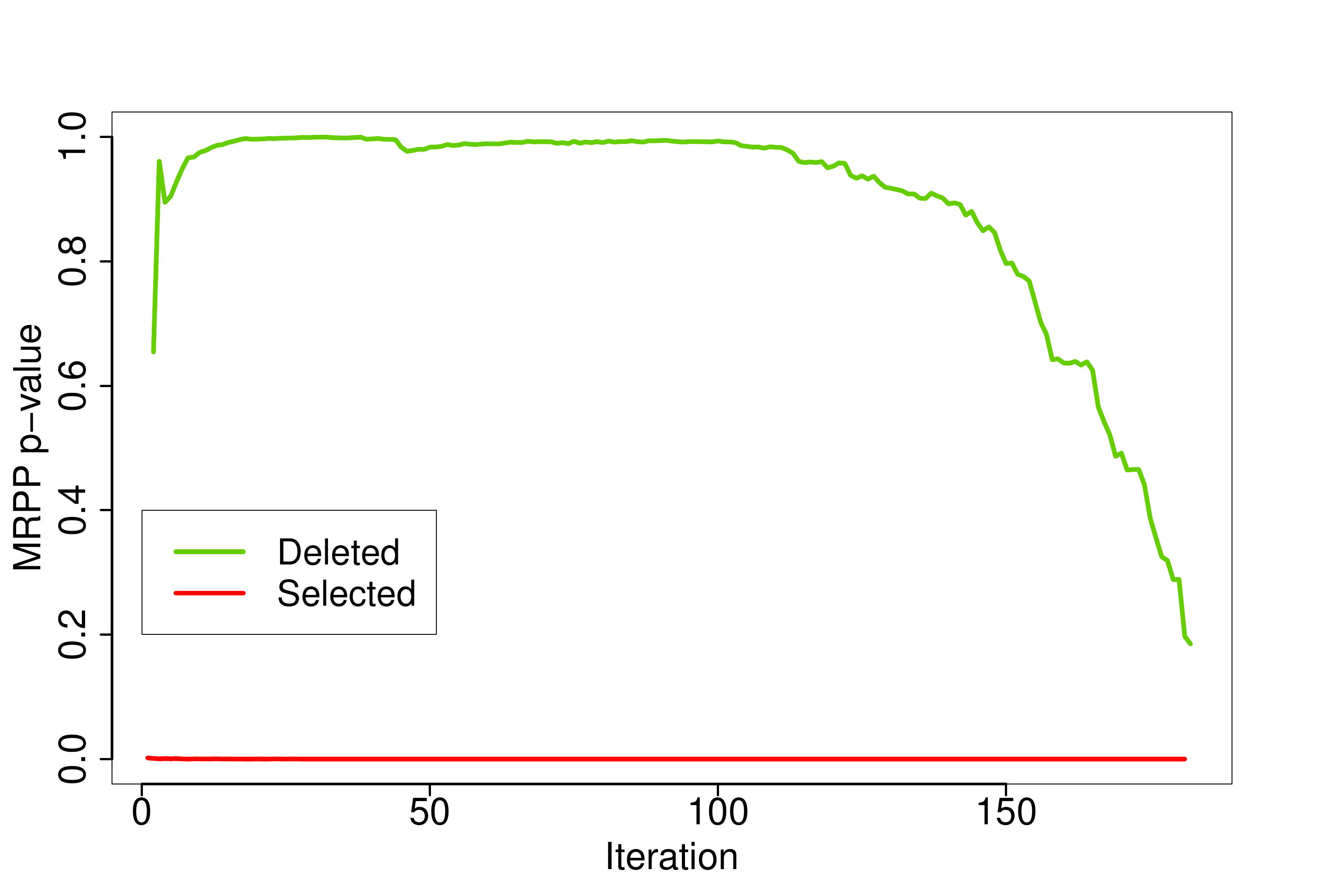}
	\caption{The MRPP $p$-values on the sets of selected and deleted genes in each iteration.}
	\label{Fig01}	
\end{figure}

We conducted backward selection on the $196$ transcription factor activity genes. The algorithm terminates after $182$ iterations ($L=182$) at which point the importance measure of each of the remaining genes is negative. The final inclusion set $\bS(L)$ contains $14$ genes. When we apply the MRPP to the $14$ selected genes, the MRPP $p$-value is less than $0.001$. On the other hand, the MRPP $p$-value on the $182$ excluded genes is $0.173$. Hence, our proposed backward selection algorithm is able to remove a majority of genes whose joint distribution does not appear to differ across groups. This allows us to focus follow-up efforts on the subset of $14$ genes judged to be important by our procedure. Figure \ref{Fig01} gives the MRPP $p$-value for the sets of selected and deleted genes in each iteration. We can see from the plot that the MRPP $p$-values on the selected genes across all iterations remains significantly small while the MRPP $p$-values for the set of deleted genes is high for the first $100$ iterations and then steadily decreases until the backward selection procedure terminates.

Form each iteration, we can collect the sign of the importance measure for each remaining gene. By considering the signs for each gene across all iterations, we obtain the proportion of iterations that each particular gene is declared as important. First, all $14$ genes in the inclusion set $\bS(L)$, have negative importance measures for all $182$ iterations. Second, if we choose $\delta=0.9$, $0.95$ and $0.99$, then the corresponding sets of important genes determined by $\bS_{\Sign}(L, \delta)$ have $20$, $18$ and $14$ genes, respectively. Thus, $\bS_{\Sign}(L, 0.99)$ is exactly the same as $\bS(L)$. Similarly, we can use the rank vectors $\left\{\Rank(\ell),~\ell=1,\ldots,L\right\}$ to rank all genes in terms of their importance in differentially expression. From the average rank of each gene, one gene not included in $\bS(L)$ emerges as the ninth most important gene, but overall importance rankings based on $\bS(L)$, $\left\{\Sign(\ell),~\ell=1,\ldots,L\right\}$ and $\left\{\Rank(\ell),~\ell=1,\ldots,L\right\}$ were similar.

There are other analysis options for identifying which of the $196$ transcription factor activity genes are most relevant to the difference between groups 1 and 2. Perhaps the most obvious approach would be to conduct a two-sample $t$-test separately for each gene.  When controlling the false discovery rate at approximately 0.05 using the method of \cite{Benjamini1995},  $11$ out of the $196$ genes are identified as differentially expressed.  A similar approach involves conducting a moderated two-sample $t$-test as implemented in the R package {\em limma} introduced in \citet{Smyth2004} and again controlling false discovery rate at approximately the 0.05 level.  This approach, which borrows information across genes to estimate the error variance for each gene, yields $12$ genes that include the $11$ identified by the traditional two-sample $t$-test approach.  The $14$ genes identified by our backward selection algorithm include the $12$ genes identified by the moderated $t$-test approach.  Thus, in this example our backward selection procedure provides some additional discoveries but overall performs similarly to the conventional approaches.

To gain further insight into the performance of the backward selection algorithm, we compare the differences between the correlation matrices of the selected genes for each treatment. Genes are usually regulated together to carry out their functions, and the proposed importance measure and backward selection method can take differences between the covariance matrices into consideration. Hence, a good subset of selected differential expressed genes may include genes whose correlation matrices differ between two groups. For the data set we analyzed, the average absolute differences between the sample correlation matrices was $0.187$ for the original $196$ genes, $0.241$ for the $14$ selected genes, and $0.182$ for the $182$ excluded genes. Thus, it seems that our procedure succeeded in focusing attention on a subset of transcription factor activity genes whose correlation structure differs across groups to a greater extent than the correlation structure of transcription factor activity genes in general.

\section{Simulations} \label{Sec_Simu}

\subsection{Backward Selection}

The results from Section \ref{Sec_Real_data} indicate that our proposed backward selection algorithm performs similarly to conventional approaches but may also possess some advantages for detecting differentially expressed genes. Because the true differential expression status of genes is unknown in applications, in this section, we examine the performance of backward selection relative to the $t$-test approaches when applied to data simulated from the ALL dataset in such a way that true differential expression status is known.

We continue to focus on the $40$ samples with BCR/ABL mutation (group 1) contrasted against the $35$ individuals without BCR/ABL mutation (group 2). Among the set of $2149$ genes that remained after applying the filtering criteria described in Section \ref{Sec_Real_data}, there are $14697$ different GO terms associated with at least one of these genes. Thus, we can define $14697$ different gene sets corresponding to these $14697$ GO terms. We focus on the subset of these $14697$ gene sets with cardinality no smaller than $40$ and MRPP $p$-value less than $ 0.05 $ for testing equality of joint expression distributions between groups 1 and 2. This results in $859$ gene sets selected for further study. The number of genes in these sets range from $40$ to $2103$. 
For each of the $859$ sets, the following procedures was used to simulate datasets.


\noindent\textbf{1}. Use the two-sample $t$-test with FDR control at $0.05$ to obtain a set of genes that are detected as differentially expressed. Denote the number of genes selected as $p_0$.

\noindent\textbf{2}.  Use {\em limma} and our backward selection algorithm to select the $p_0$ most significant genes according to each method. Find the union of the three sets obtained from two-sample $t$-test,  {\em limma} and backward selection, denoted as $\Theta$, and let $p_1$ be the number of genes in $\Theta$.

\noindent\textbf{3}.  Randomly select $2\times 15$ individuals without replacement from group 1, and randomly divide the selected samples into two groups, each with $15$ samples. Denote these two groups as group $1'$ and group $2'$.

\noindent\textbf{4}. 
For each individual in group $2'$ created in step 3, replace the data for the $p_1$ genes in $\Theta$ with 
data for the $p_1$ genes in $\Theta$ from $15$ samples selected without replacement from group 2.

\begin{figure}[H]
	\centering
	\subfigure[$\rho_t-\rho_{bs}$]{%
		\includegraphics[width=0.3\linewidth]{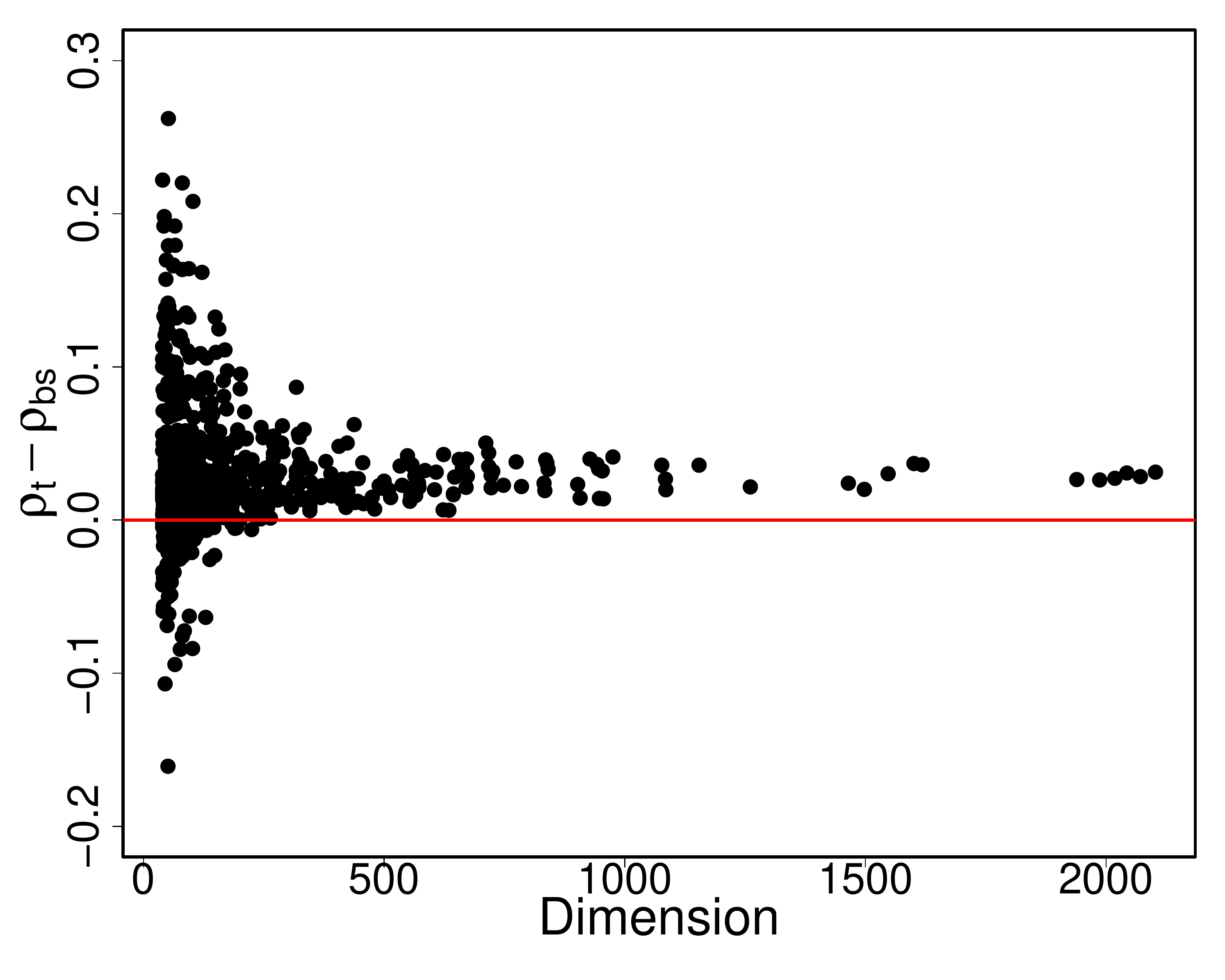}
		\label{fig:t_bs}}
	\quad
	\subfigure[$\rho_l-\rho_{bs}$]{%
		\includegraphics[width=0.3\linewidth]{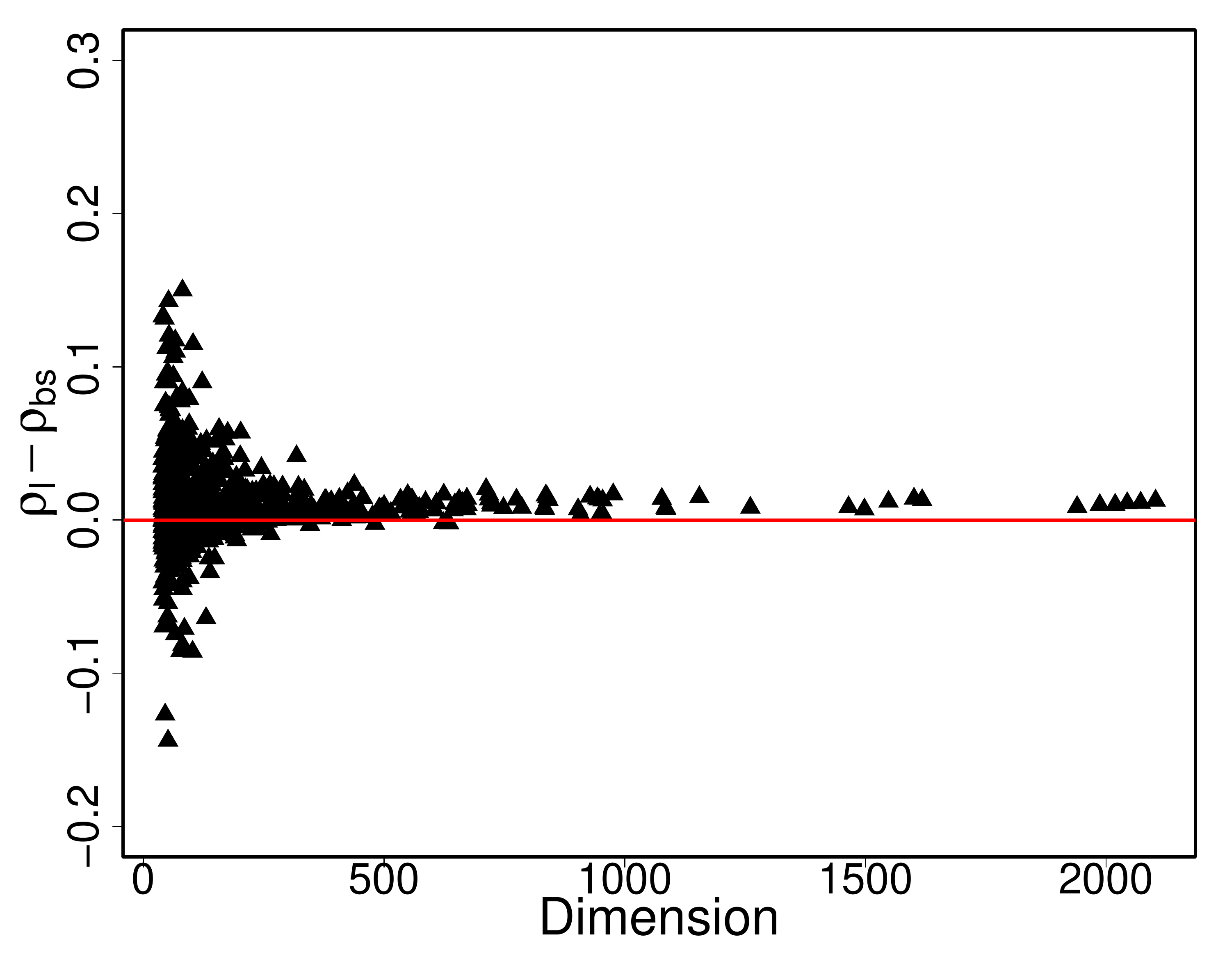}
		\label{fig:limma_bs}}
	\quad
	\subfigure[$\rho_t-\rho_{l}$]{%
		\includegraphics[width=0.3\linewidth]{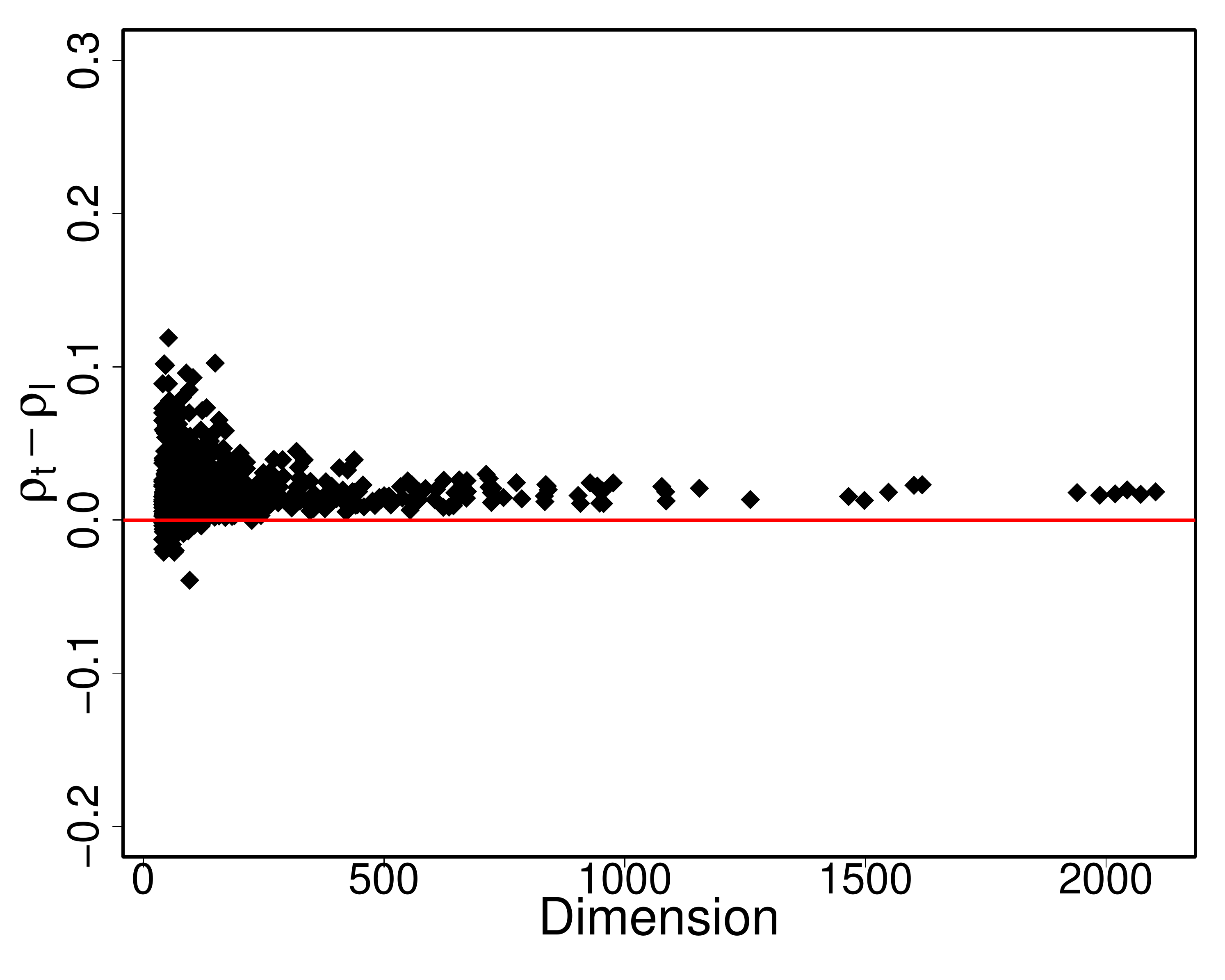}
		\label{fig:t_limma}}
	\quad	
	\caption{Comparison of average false positive rate for three methods for each gene set, $\rho_t$ for two-sample $t$-test, $\rho_l$ for {\em limma} and $\rho_{bs}$ for backward selection. The horizontal axis represents the dimension of each gene set and the vertical axis gives: (a) $\rho_t-\rho_{bs}$; (b) $\rho_l-\rho_{bs}$; (c) $\rho_t-\rho_{l}$. 
	}
	\label{fig:ALL_simu_summ}
\end{figure}

These four steps produce two samples, each of size $15$, drawn from multivariate distributions that differ only for the subset of dimensions corresponding to $\Theta$. 
Following steps $1$ through $4$, we simulate $1000$ data sets for each gene set. For each simulated set, we find the $p_1$ most significant genes using the two-sample $t$-test, {\em limma}, and our backward selection procedure. For the backward selection algorithm in the simulation, the most significant genes are determined according to the average of rank vectors $\{\Rank(\ell), \ell = 1, \cdots, L\}$. We then calculate the percentage of the selected genes not in $\Theta$ for each of the three methods and average the results over $1000$ simulation replications to get the average false positive rate for the three methods. Let $\rho_t$, $\rho_{l}$ and $\rho_{bs}$ represent those average false positive rates for the two-sample $t$-test, {\em limma}, and backward selection, respectively. The results for comparing the three methods are given in Figure \ref{fig:ALL_simu_summ}.

By looking at plots (a) through (c) in Figure \ref{fig:ALL_simu_summ}, we see that our backward selection approach tends to produce the lowest false positive rates (and thus the highest discovery rates), especially for high-dimensional gene sets. The two-sample $t$-test has worst performance against the other two methods. 
In addition, Table \ref{Tab01} gives the number of gene sets for which each method ranks first, second, or third among the three methods with respect to false positive rate among the $859$ gene sets. 
Our backward selection algorithm has the 
lowest false positive rate among the three methods for nearly $70\%$ of the gene sets.
\begin{table}[ht]
	\caption{Number of gene sets for which each method ranks $1$, $1.5$, $2$, $2.5$ or $3$ with respect to false positive rate among the $859$ gene sets. (Rank $1$ is best and $3$ is worst, $1.5$ and $2.5$ correspond to ties.)}
	\centering
	\begin{tabular}{cccccc}
		\hline 
		& \multicolumn{5}{c}{Rank}\\
		\cline{2-6}
		Method & 1 & 1.5 & 2 & 2.5 & 3 \\
		\hline
		Two-sample $t$-test & 27 & 1 & 129 & 4 & 698 \\
		{\em limma} & 210 & 6 & 577 & 0 & 66 \\
		Backward selection & 616 & 5 & 143 & 4 & 91 \\
		\hline
	\end{tabular}
	\label{Tab01}
\end{table}

\begin{figure}[h!]
	\centering
	\subfigure[$\rho_t-\rho_{bs}$]{%
		\includegraphics[width=0.3\linewidth]{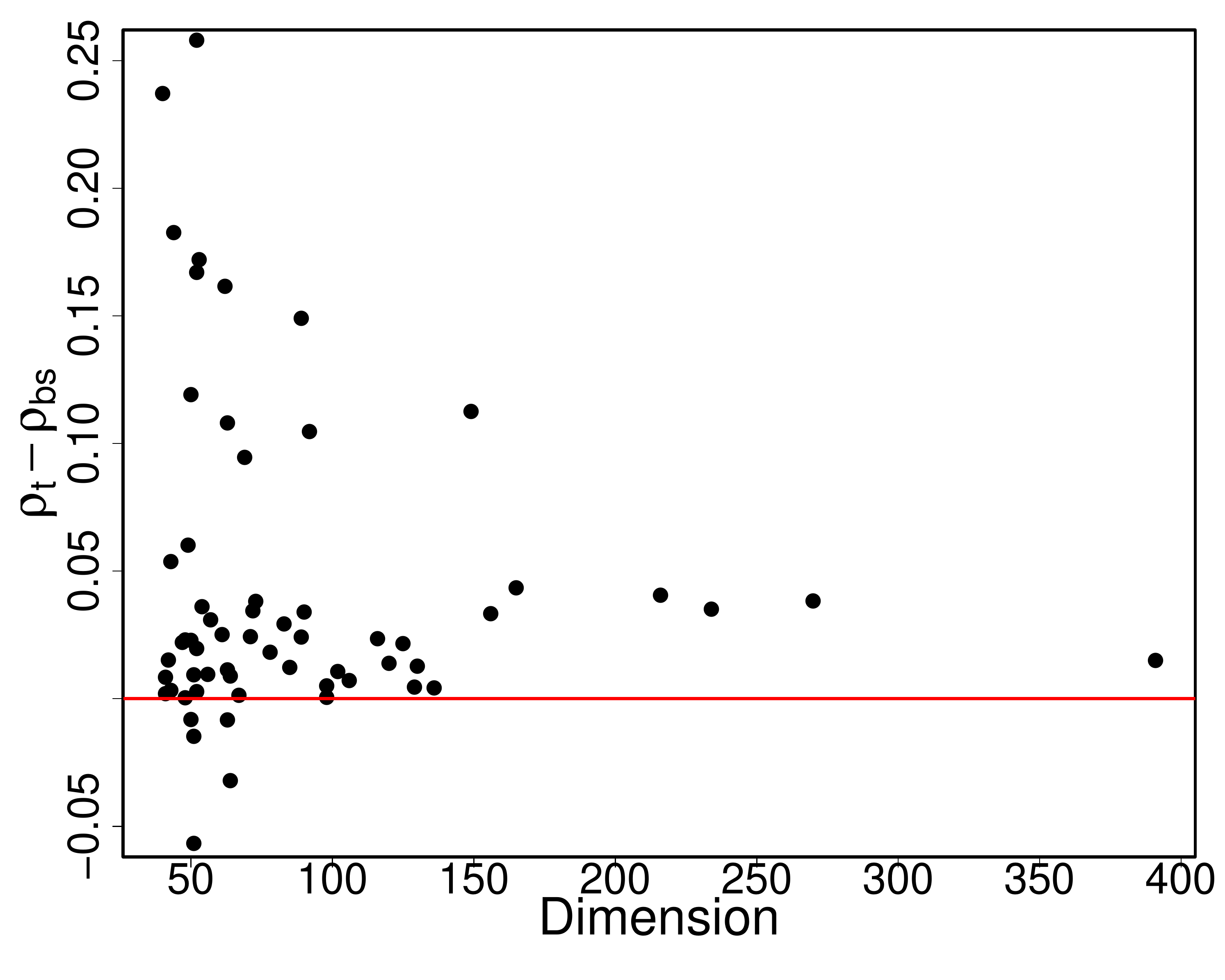}
		\label{fig:t_bs1}}
	\quad
	\subfigure[$\rho_l-\rho_{bs}$]{%
		\includegraphics[width=0.3\linewidth]{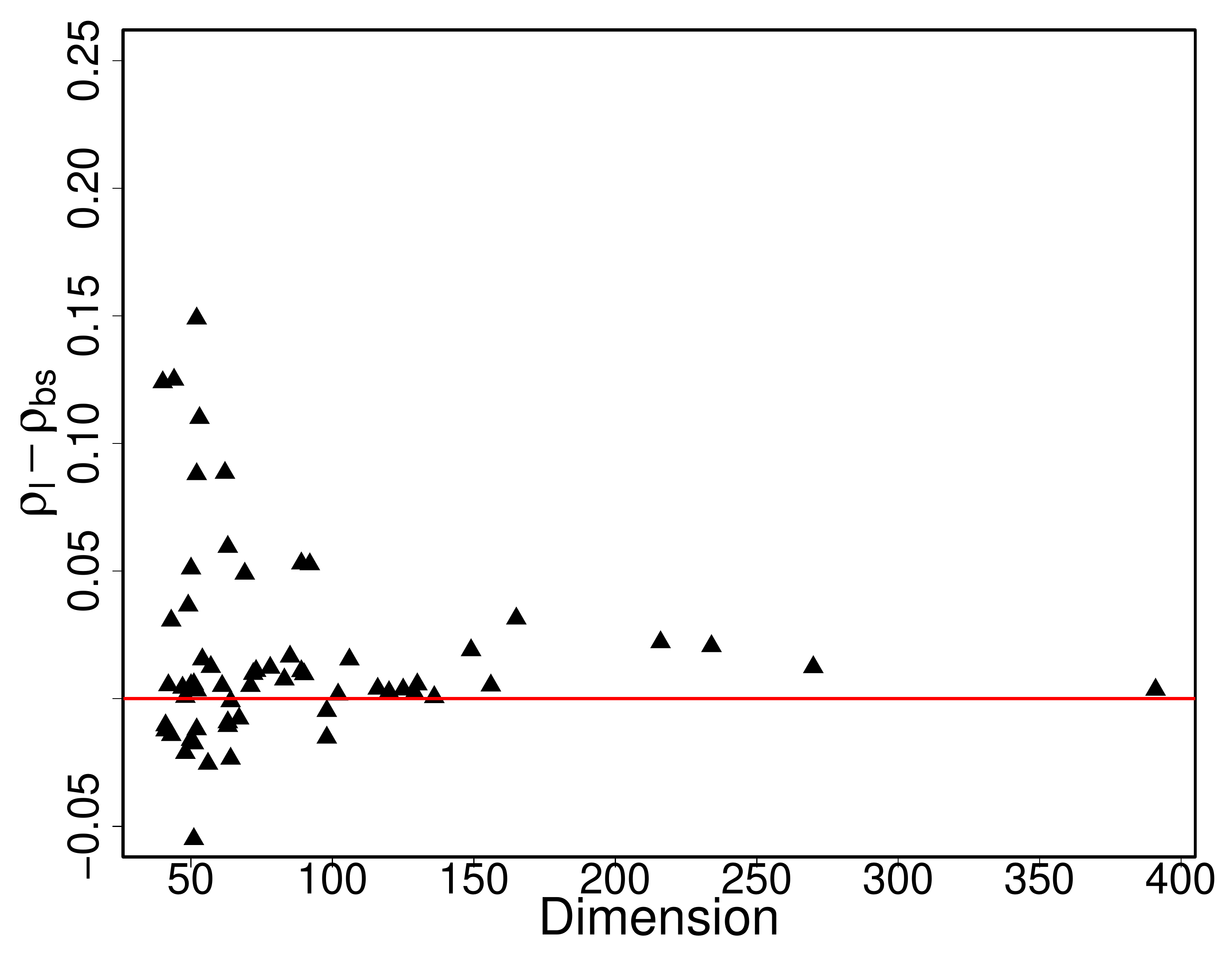}
		\label{fig:limma_bs1}}
	\quad
	\subfigure[$\rho_t-\rho_{l}$]{%
		\includegraphics[width=0.3\linewidth]{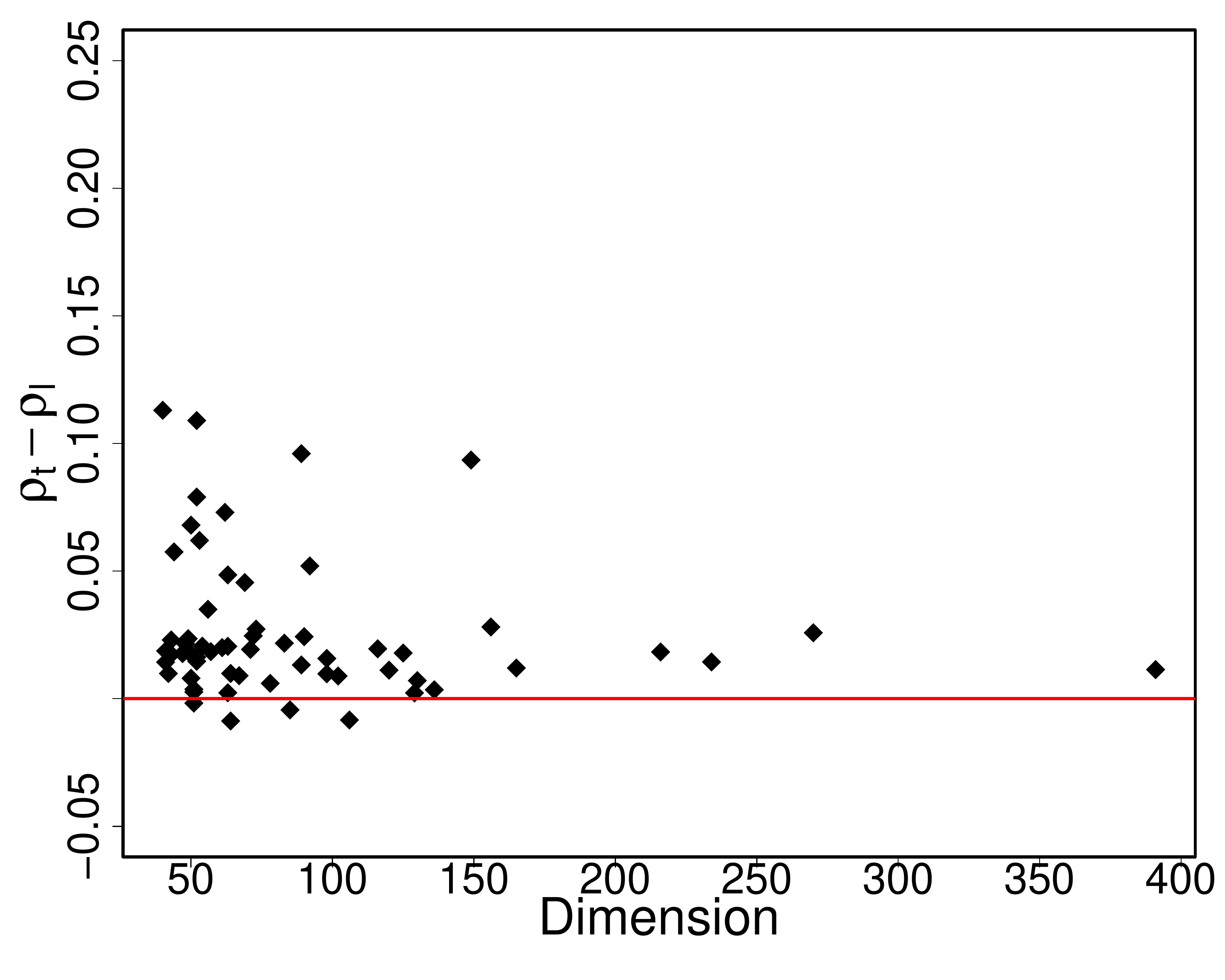}
		\label{fig:t_limma1}}
	\quad	
	\caption{Comparison of average false positive rate for three methods when applied to the $61$ gene sets, $\rho_t$ for two-sample $t$-test, $\rho_l$ for {\em limma} and $\rho_{bs}$ for backward selection. The x-axis represents the dimension of each gene set and the y-axis gives: (a) $\rho_t-\rho_{bs}$; (b) $\rho_l-\rho_{bs}$; (c) $\rho_t-\rho_{l}$.}
	\label{fig:ALL_simu_summ1}
\end{figure}

In the previous simulation, the number of selected genes for each method is determined by the two-sample $t$-test, which can potentially produce bias. To avoid this, instead of following step 1 and 2 when simulating datasets, we use the two-sample $t$-test with FDR control at $0.05$, {\em limma} with FDR control at $0.05$ and our proposed backward selection algorithm with inclusion set $\mathbf{S}(L)$ to obtain three sets of genes that are detected as differentially expressed. We then focus on the gene sets in which the number of genes selected was the same for all three methods, which results in $61$ gene sets. The number of genes in those $61$ gene sets ranges from $40$ to $391$. For each of the $61$ gene sets, we let $\Theta$ be the union of the three gene sets obtained from the two-sample $t$-test,  {\em limma} and backward selection, and let $p_1$ be the number of genes in $\Theta$. We then follow the same simulation steps described previously. The results for the $61$ gene sets are shown in Figure \ref{fig:ALL_simu_summ1}.


From Figure \ref{fig:ALL_simu_summ1}, we see that for most of the gene sets, the backward selection algorithm performs best and that {\em limma} performs better than the traditional two-sample $t$-test. Table \ref{Tab02} presents the number of gene sets for which each method ranks first, second, or third among the three methods with respect to false positive rate. 
Based on our simulation results, we conclude that, overall, the proposed backward selection method performs best among the three methods, and its performance advantage increases in high-dimensional situations.

\begin{table}[ht]
	\caption{Number of gene sets for which each method ranks $1$, $2$, or $3$ with respect to false positive rate among the $61$ gene sets. (Rank $1$ is best and $3$ is worst, no ties found in the results.)}
	\centering
	\begin{tabular}{cccc}
		\hline 
		& \multicolumn{3}{c}{Rank}\\
		\cline{2-4}
		Method & 1 & 2 & 3 \\
		\hline
		Two-sample $t$-test & 2 & 5 & 54 \\
		{\em limma} & 14 & 45 & 2 \\
		Backward selection & 45 & 11 & 5 \\
		\hline
	\end{tabular}
	\label{Tab02}
\end{table}

\subsection{Modified MRPP}

A modified MRPP test procedure based on our proposed backward selection algorithm is introduced in Section \ref{sec:modified_MRPP}. In this section, we compare the performance of our proposed modified MRPP with the original MRPP for testing differences between two multivariate distributions by conducting Monte Carlo simulations. We consider six different sample pairs $(n_1, n_2)$ in combination five different choices for the data vector dimension $R$. We focus primarily on combinations where the dimension exceeds the sample size (see Table \ref{tab:n_1_n_2}). All our size and power estimates are based on $1000$ Monte Carlo simulations at the nominal level $\alpha=0.05$.

\begin{table}[h]
	\centering
	\caption{Combinations of $n_1$ and $n_2$ considered in the simulations.}
	\begin{tabular}{c|cccccc}
		\hline
		&  1 &  2 &  3 &  4 &  5 & 6\\\hline
		$n_1$ & 20 & 20 & 20 & 40 & 40 & 80\\ 
		$n_2$ & 20 & 40 & 80 & 40 & 80 & 80\\\hline
		$N$ & 40 & 60 & 100 & 80 & 120 & 160 \\\hline
	\end{tabular}\label{tab:n_1_n_2}
\end{table}

First, we evaluate the size of the proposed test. The data $\{\bY_1, \ldots, \bY_{n_1}; \bY_{n_1+1}, \ldots, \bY_N\}$ are generated from the $R$-dimensional multivariate normal distribution with mean vector $\mathbf{0}_R$ and covariance matrix $\bSigma=\left(0.5^{|i-j|}\right)_{1\leq i,j\leq R}$. For each simulated data set, we carry out the original MRPP along with our proposed modified MRPP for testing for distributional differences between two groups $\{\bY_1, \ldots, \bY_{n_1}\}$ and $\{\bY_{n_1+1}, \ldots, \bY_N\}$. The number of permutations is set at $1000$ for both testing procedures. To implement the modified MRPP, we choose the number of variables selected ($R_0$) in several different ways.
First, we consider setting $R_0$ to be the number of variables in the inclusion set $\bS(L)$ obtained from backward selection. Alternative choices are obtained by prespecifying $R_0$ as $2$, $4$, $8$, $16$, or $\sqrt{R}$.

The sizes for both original and our modified MRPP tests summarized in Table \ref{tab:size} show that the original MRPP test maintains the size well around the nominal significant level $0.05$. When the number of variables $R_0$ is pre-specified before the testing procedure, the sizes of the modified MRPP can also be well controlled. However, when $R_0$ is chosen as the cardinality of the inclusion set $\bS(L)$ after the backward selection procedure, the sizes of the modified MRPP are slightly larger than the nominal level. Overall, the modified MRPP has good control of Type-I error under the null hypothesis that the distributions of two groups are the same.

\begin{table}[h!]
	\renewcommand{\arraystretch}{1}
	\centering
	\caption{Empirical sizes of original MRPP ($\mathrm{MRPP}_{Org}$) and modified MRPP with different choices of $R_0$. For $\Mod_{\bS(L)}$, $R_0$ is chosen as the cardinality of $\bS(L)$. For $\Mod_2$, $\Mod_4$, $\Mod_8$, $\Mod_{16}$ and $\Mod_{\sqrt{R}}$, $R_0$ is pre-specified as $2$, $4$, $8$, $16$ and $\sqrt{R}$, respectively.}
	\begin{tabular}{c|ccccccccccc}
		\hline
		$R$ & 25 & 50 & 100 & 200 & 400 & & 25 & 50 & 100 & 200 & 400 \\\hline
		& \multicolumn{5}{c}{$n_1=20,~n_2=20$} & & \multicolumn{5}{c}{$n_1=20,~n_2=40$} \\
		$\mathrm{MRPP}_{Org}$ & 0.042 & 0.047 & 0.047 & 0.056 & 0.044 & & 0.052 & 0.055 & 0.052 & 0.044 & 0.059 \\
		$\Mod_{\bS(L)}$ & 0.064 & 0.074 & 0.063 & 0.063 & 0.050 & & 0.075 & 0.068 & 0.073 & 0.056 & 0.055 \\
		$\Mod_2$ & 0.046 & 0.046 & 0.051 & 0.047 & 0.046 & & 0.046 & 0.041 & 0.058 & 0.046 & 0.046 \\
		$\Mod_{4}$ & 0.038 & 0.047 & 0.045 & 0.046 & 0.044 & & 0.050 & 0.044 & 0.052 & 0.043 & 0.041 \\
		$\Mod_{8}$ & 0.038 & 0.042 & 0.044 & 0.053 & 0.042 & & 0.055 & 0.044 & 0.053 & 0.042 & 0.042 \\
		$\Mod_{16}$ & 0.042 & 0.040 & 0.042 & 0.045 & 0.040 & & 0.053 & 0.056 & 0.052 & 0.035 & 0.043 \\
		$\Mod_{\sqrt{R}}$ & 0.043 & 0.047 & 0.042 & 0.047 & 0.042 & & 0.049 & 0.042 & 0.052 & 0.036 & 0.044 \\\hline
		& \multicolumn{5}{c}{$n_1=20,~n_2=80$} & & \multicolumn{5}{c}{$n_1=40,~n_2=40$} \\
		$\mathrm{MRPP}_{Org}$ & 0.047 & 0.045 & 0.047 & 0.049 & 0.044 & & 0.049 & 0.048 & 0.052 & 0.041 & 0.046 \\
		$\Mod_{\bS(L)}$ & 0.063 & 0.063 & 0.055 & 0.065 & 0.056 & & 0.070 & 0.065 & 0.061 & 0.063 & 0.044 \\
		$\Mod_2$ & 0.044 & 0.046 & 0.040 & 0.046 & 0.054 & & 0.043 & 0.044 & 0.034 & 0.051 & 0.049 \\
		$\Mod_{4}$ & 0.044 & 0.037 & 0.048 & 0.040 & 0.049 & & 0.048 & 0.039 & 0.043 & 0.050 & 0.050 \\
		$\Mod_{8}$ & 0.041 & 0.035 & 0.038 & 0.043 & 0.053 & & 0.045 & 0.041 & 0.046 & 0.054 & 0.044 \\
		$\Mod_{16}$ & 0.046 & 0.039 & 0.039 & 0.047 & 0.056 & & 0.048 & 0.041 & 0.043 & 0.055 & 0.035 \\
		$\Mod_{\sqrt{R}}$ & 0.041 & 0.038 & 0.037 & 0.044 & 0.056 & & 0.039 & 0.041 & 0.043 & 0.056 & 0.037 \\\hline
		& \multicolumn{5}{c}{$n_1=40,~n_2=80$} & & \multicolumn{5}{c}{$n_1=80,~n_2=80$} \\
		$\mathrm{MRPP}_{Org}$ & 0.038 & 0.035 & 0.042 & 0.056 & 0.046 & & 0.051 & 0.041 & 0.057 & 0.047 & 0.058 \\
		$\Mod_{\bS(L)}$ & 0.064 & 0.061 & 0.060 & 0.072 & 0.056 & & 0.071 & 0.066 & 0.071 & 0.071 & 0.057 \\
		$\Mod_2$ & 0.044 & 0.042 & 0.042 & 0.044 & 0.054 & & 0.047 & 0.044 & 0.052 & 0.048 & 0.053 \\
		$\Mod_{4}$ & 0.044 & 0.043 & 0.041 & 0.054 & 0.052 & & 0.043 & 0.044 & 0.047 & 0.055 & 0.056 \\
		$\Mod_{8}$ & 0.040 & 0.042 & 0.041 & 0.054 & 0.044 & & 0.045 & 0.039 & 0.046 & 0.051 & 0.050 \\
		$\Mod_{16}$ & 0.039 & 0.041 & 0.041 & 0.059 & 0.049 & & 0.048 & 0.039 & 0.046 & 0.058 & 0.047 \\
		$\Mod_{\sqrt{R}}$ & 0.039 & 0.040 & 0.042 & 0.056 & 0.052 & & 0.043 & 0.040 & 0.046 & 0.059 & 0.046 \\\hline			
	\end{tabular}\label{tab:size}
\end{table}

\begin{figure}[h]
	\centering
	\subfigure[$n_1=20$, $n_2=20$]{%
		\includegraphics[width=0.45\linewidth]{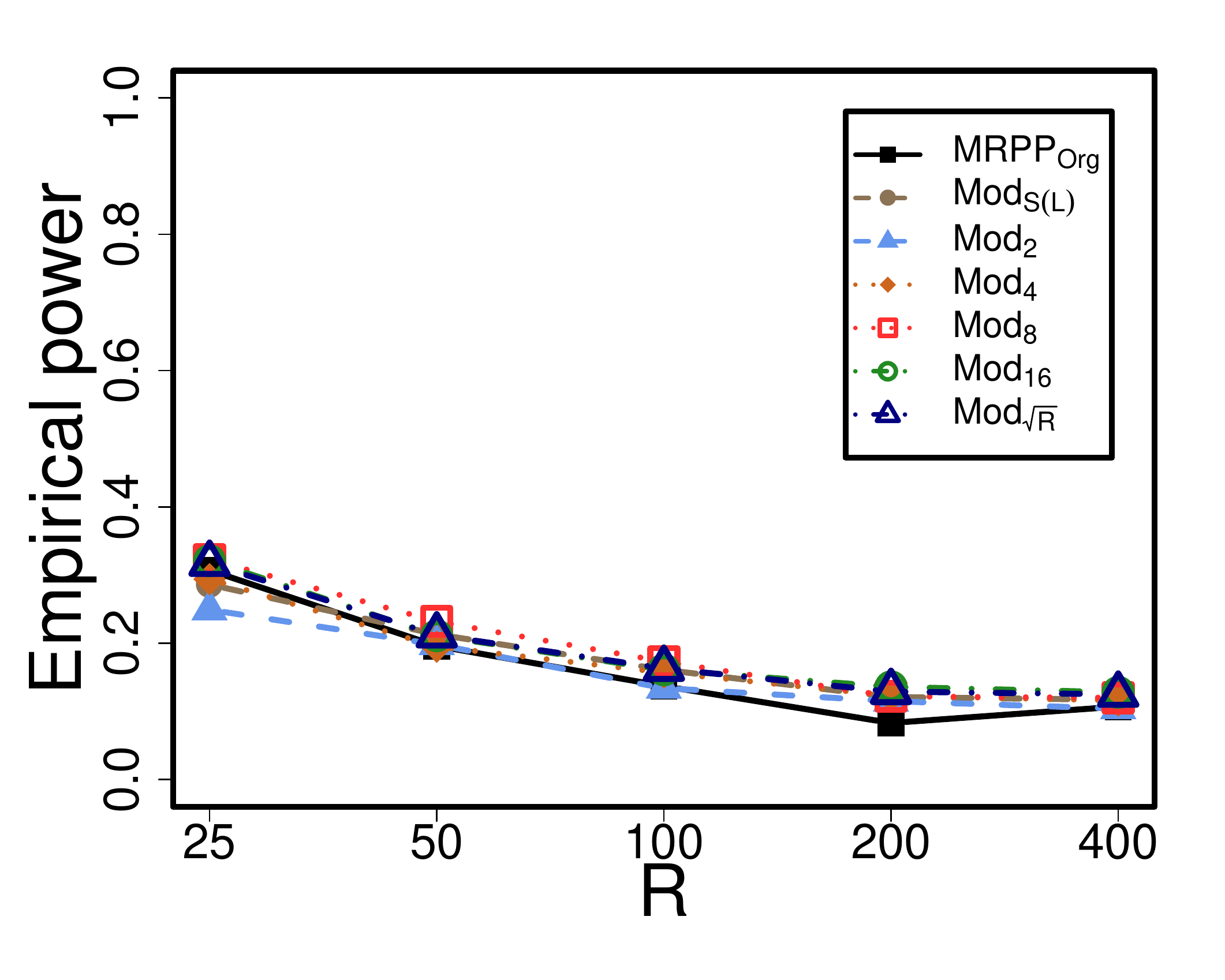}
		\label{fig:power_5_n1_20_n2_20}}
	\quad
	\subfigure[$n_1=20$, $n_2=40$]{%
		\includegraphics[width=0.45\linewidth]{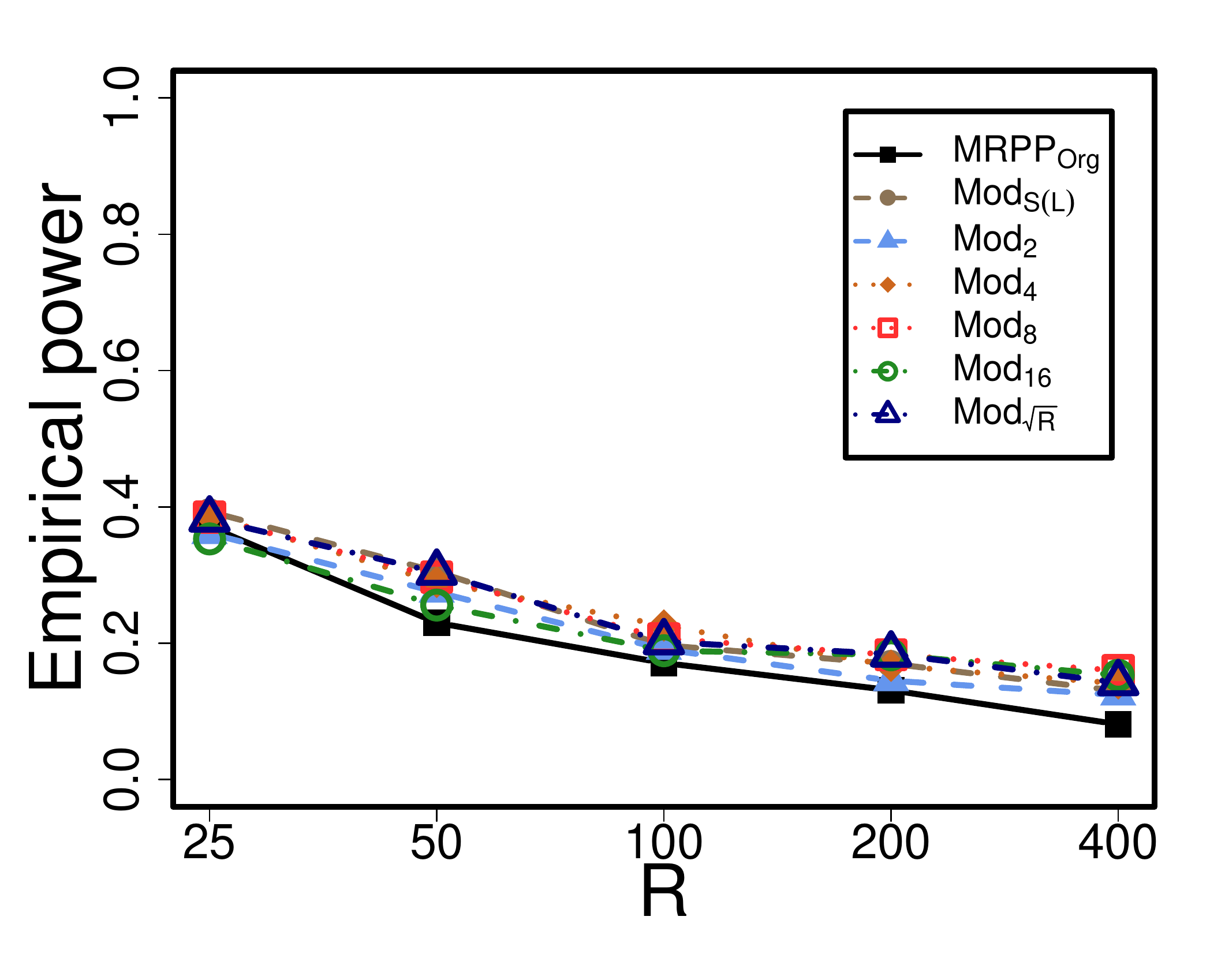}
		\label{fig:power_5_n1_20_n2_40}}
	\quad
	\subfigure[$n_1=20$, $n_2=80$]{%
		\includegraphics[width=0.45\linewidth]{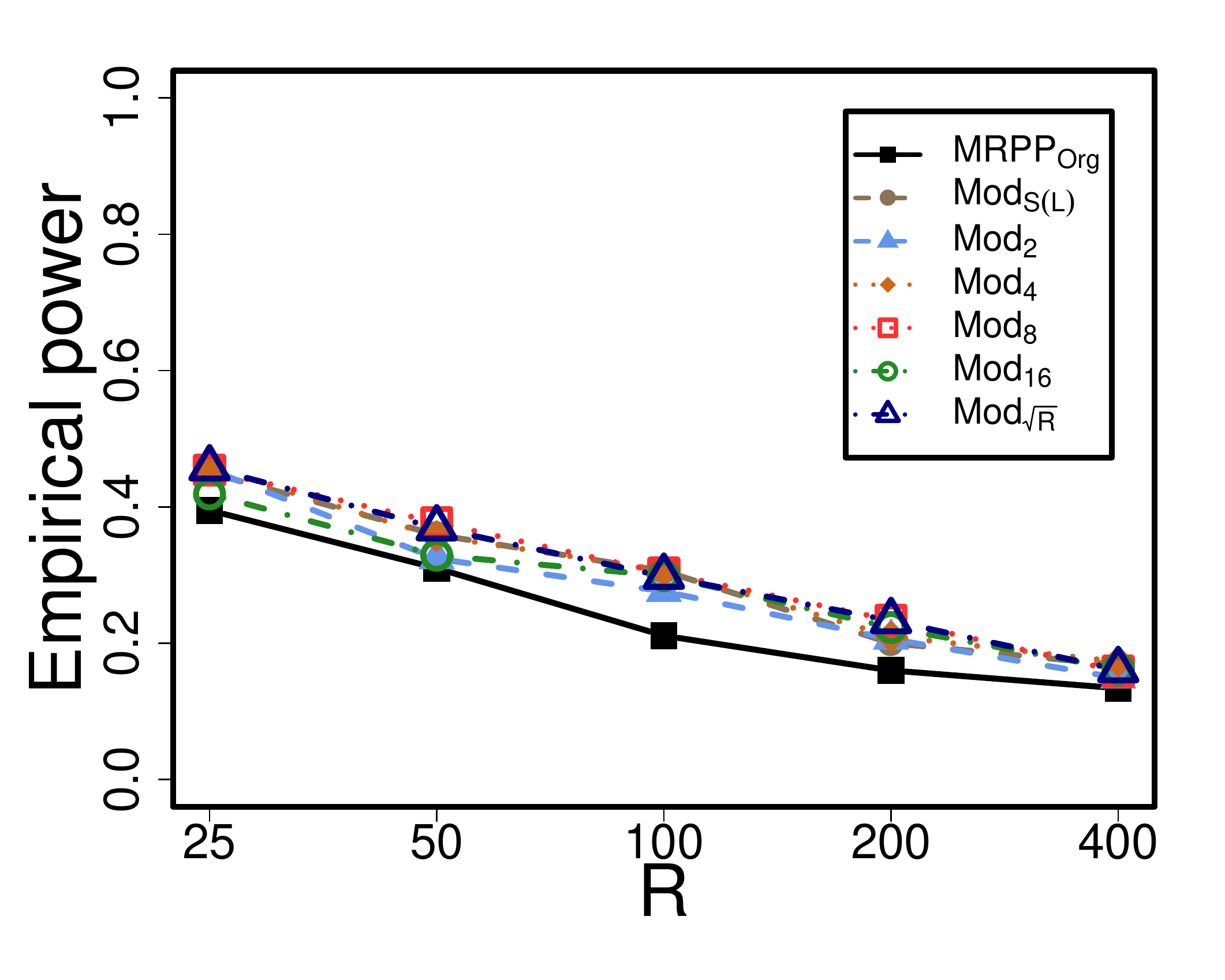}
		\label{fig:power_5_n1_20_n2_80}}
	\quad
	\subfigure[$n_1=40$, $n_2=40$]{%
		\includegraphics[width=0.45\linewidth]{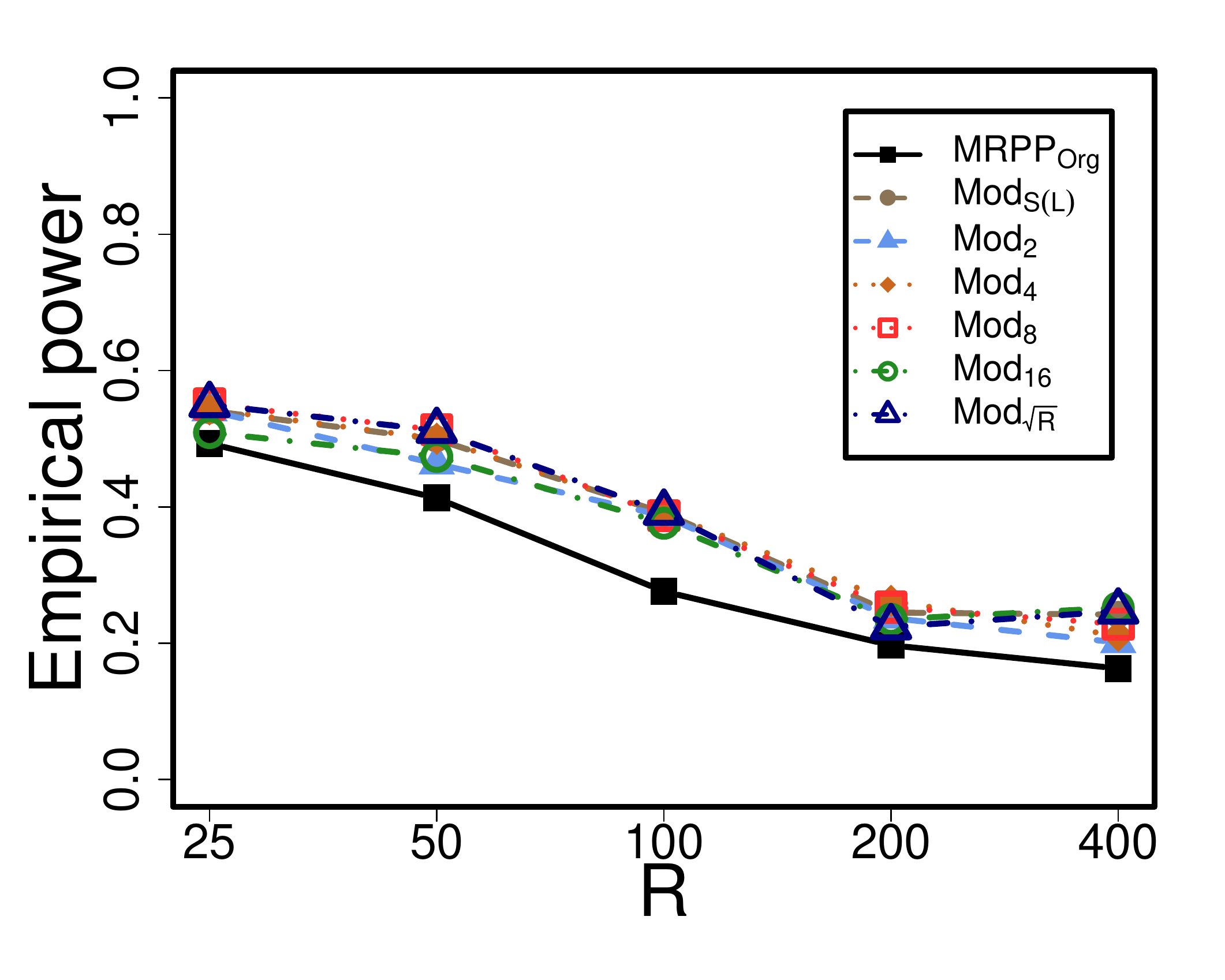}
		\label{fig:power_5_n1_40_n2_40}}
	\quad
	\subfigure[$n_1=40$, $n_2=80$]{%
		\includegraphics[width=0.45\linewidth]{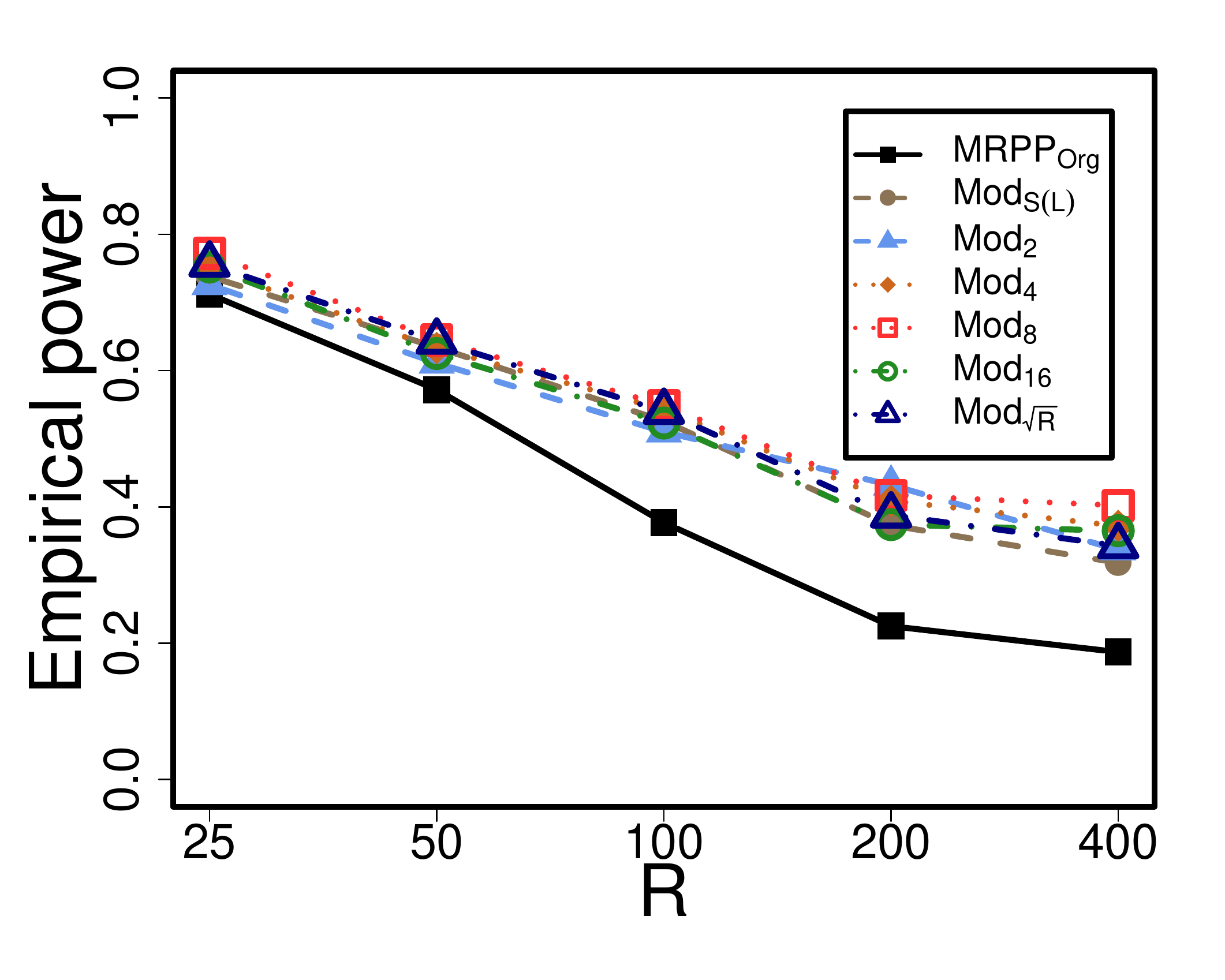}
		\label{fig:power_5_n1_40_n2_80}}
	\quad
	\subfigure[$n_1=80$, $n_2=80$]{%
		\includegraphics[width=0.45\linewidth]{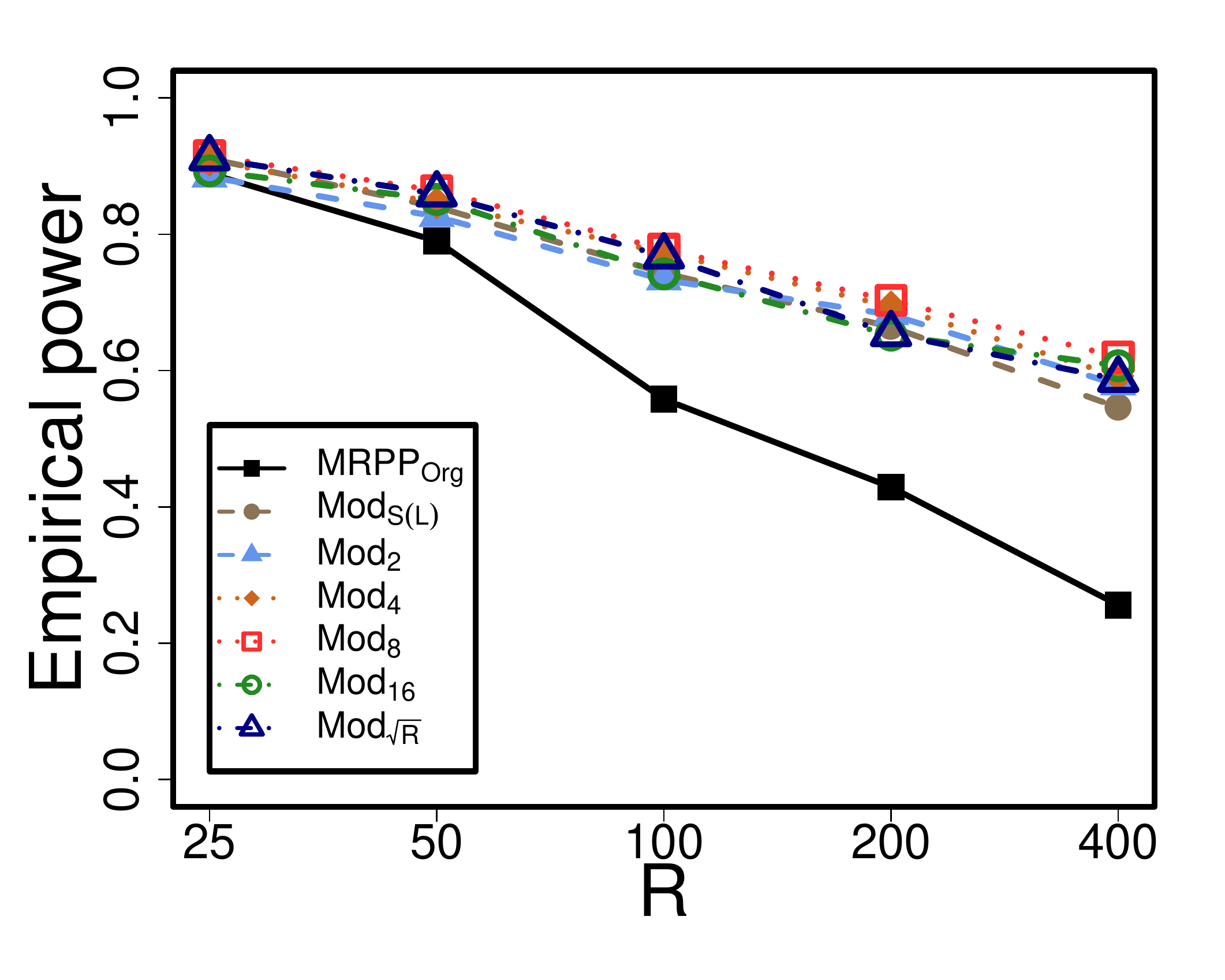}
		\label{fig:power_5_n1_80_n2_80}}
	\quad
	\caption{Empirical power of the original MRPP and modified MRPP with different choices of $R_0$ for $\nu=0.5$}
	\label{fig:modified_MRPP_power1}
\end{figure}

\begin{figure}[h]
	\centering
	\subfigure[$n_1=20$, $n_2=20$]{%
		\includegraphics[width=0.45\linewidth]{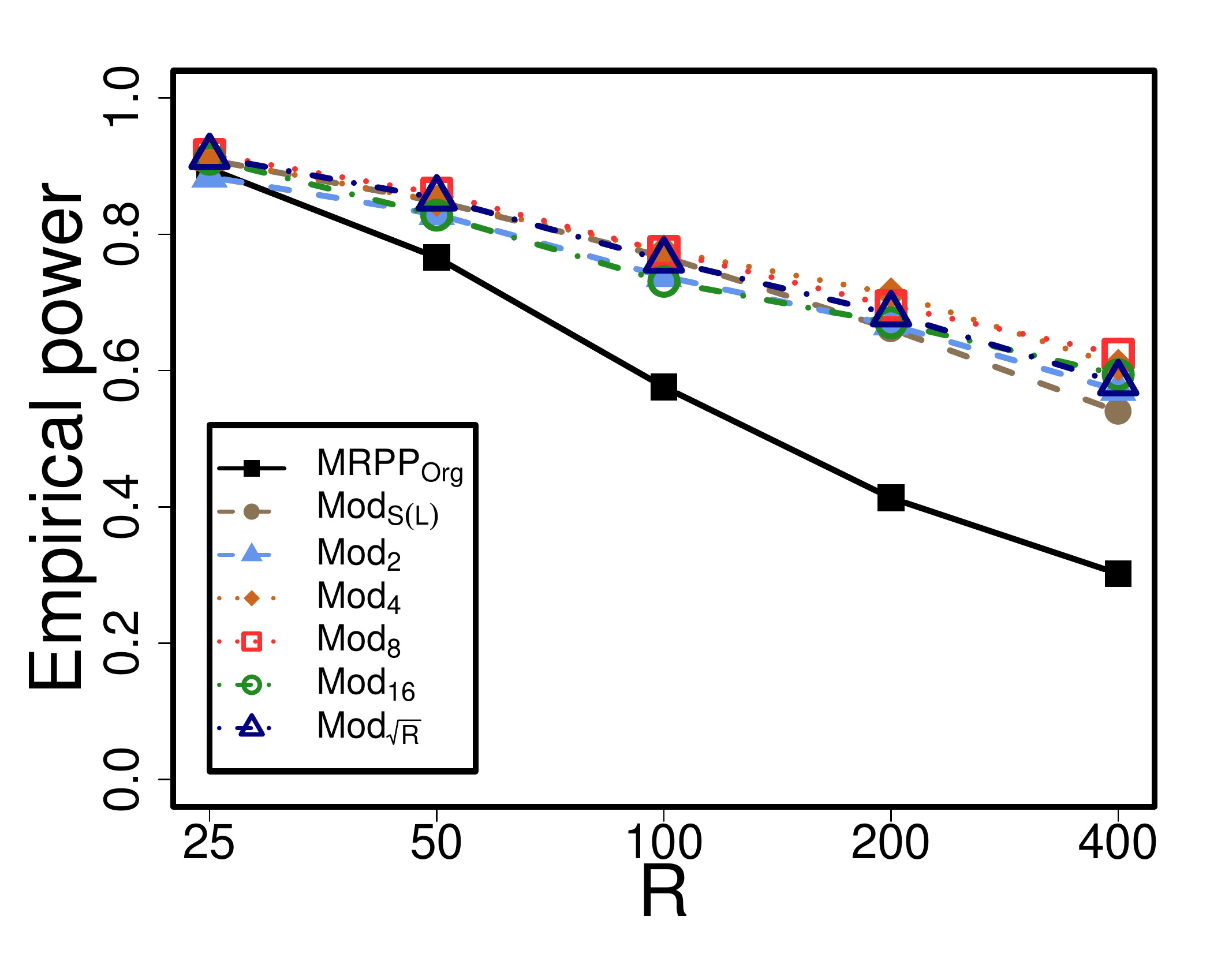}
		\label{fig:power_10_n1_20_n2_20}}
	\quad
	\subfigure[$n_1=20$, $n_2=40$]{%
		\includegraphics[width=0.45\linewidth]{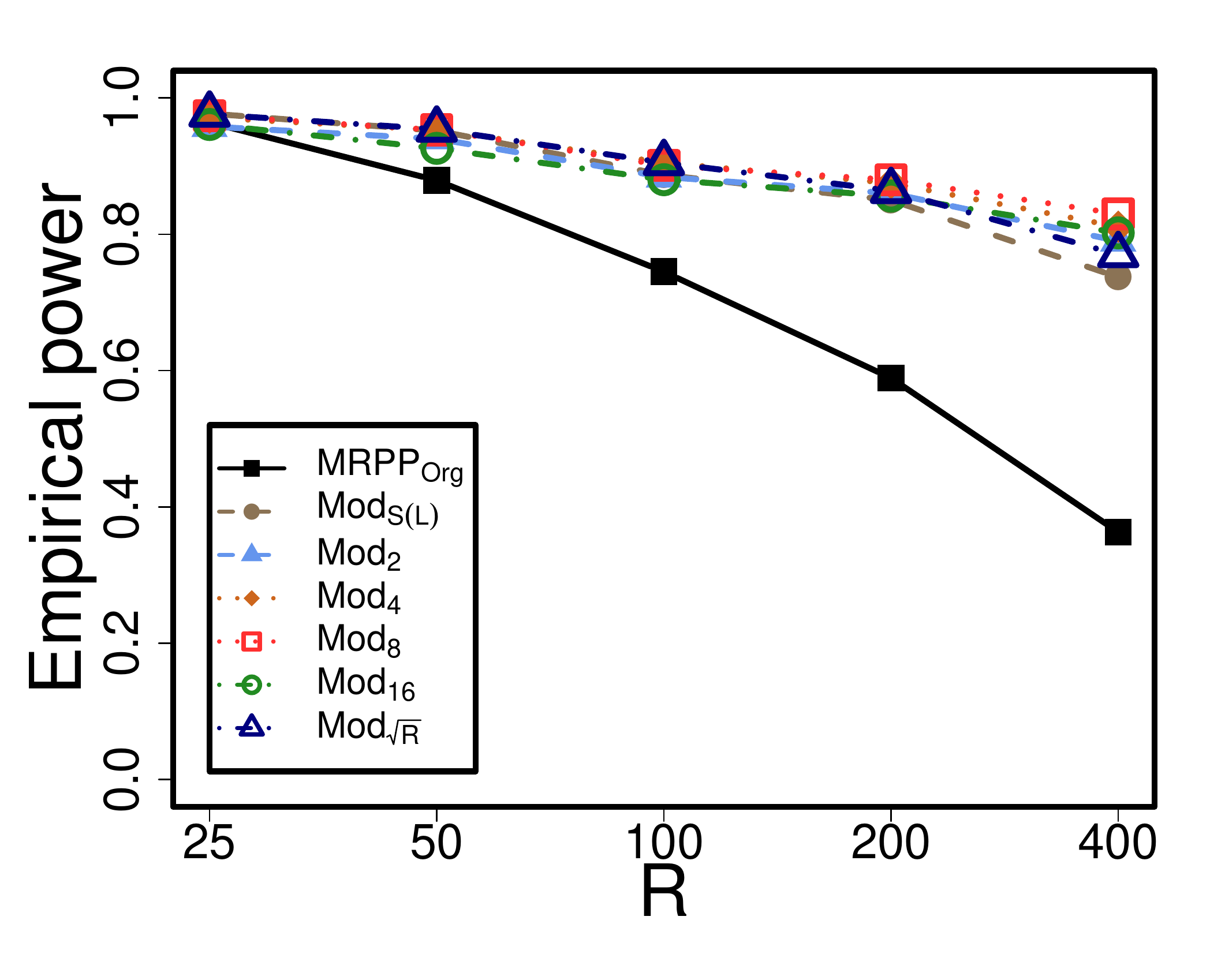}
		\label{fig:power_10_n1_20_n2_40}}
	\quad
	\subfigure[$n_1=20$, $n_2=80$]{%
		\includegraphics[width=0.45\linewidth]{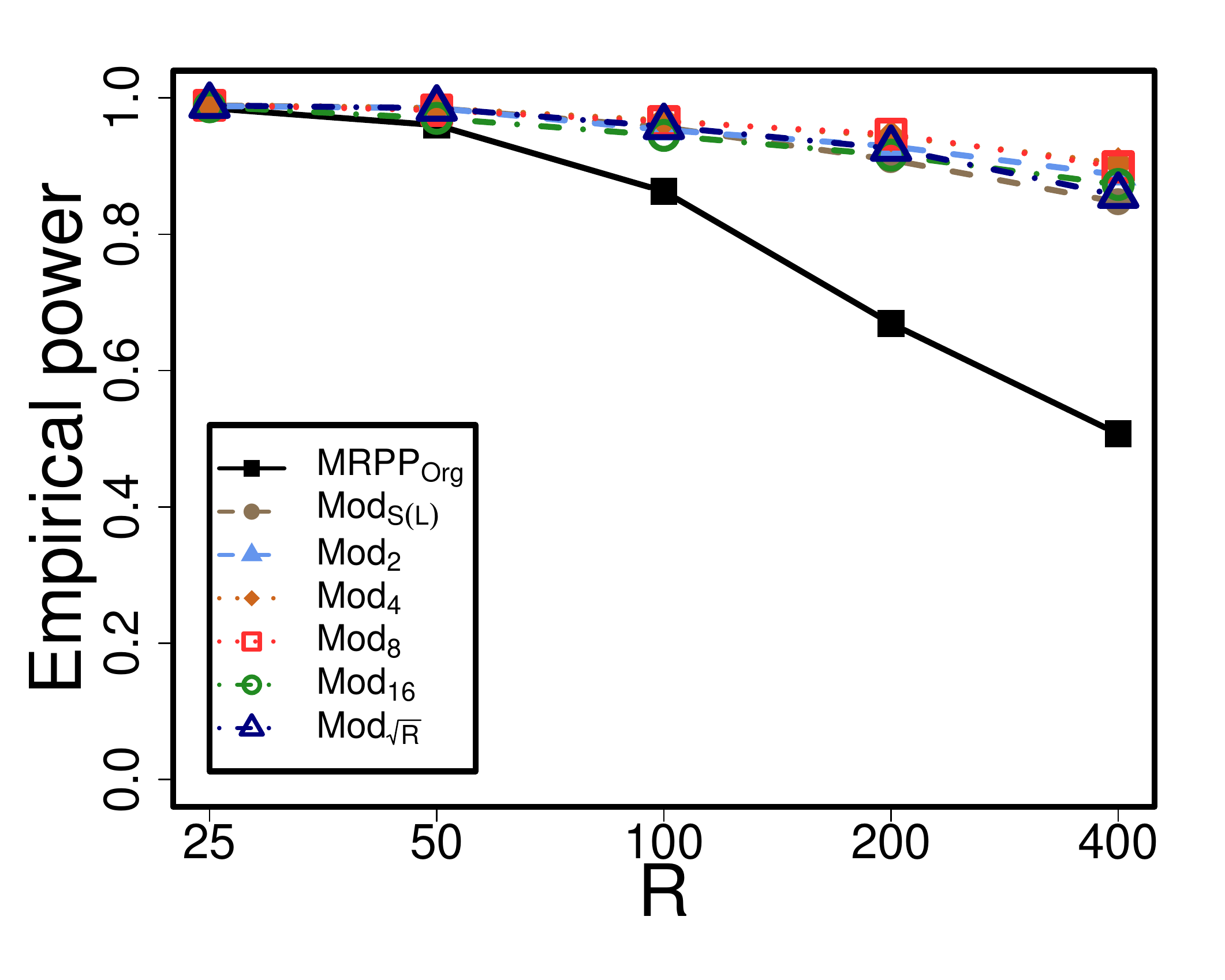}
		\label{fig:power_10_n1_20_n2_80}}
	\quad
	\subfigure[$n_1=40$, $n_2=40$]{%
		\includegraphics[width=0.45\linewidth]{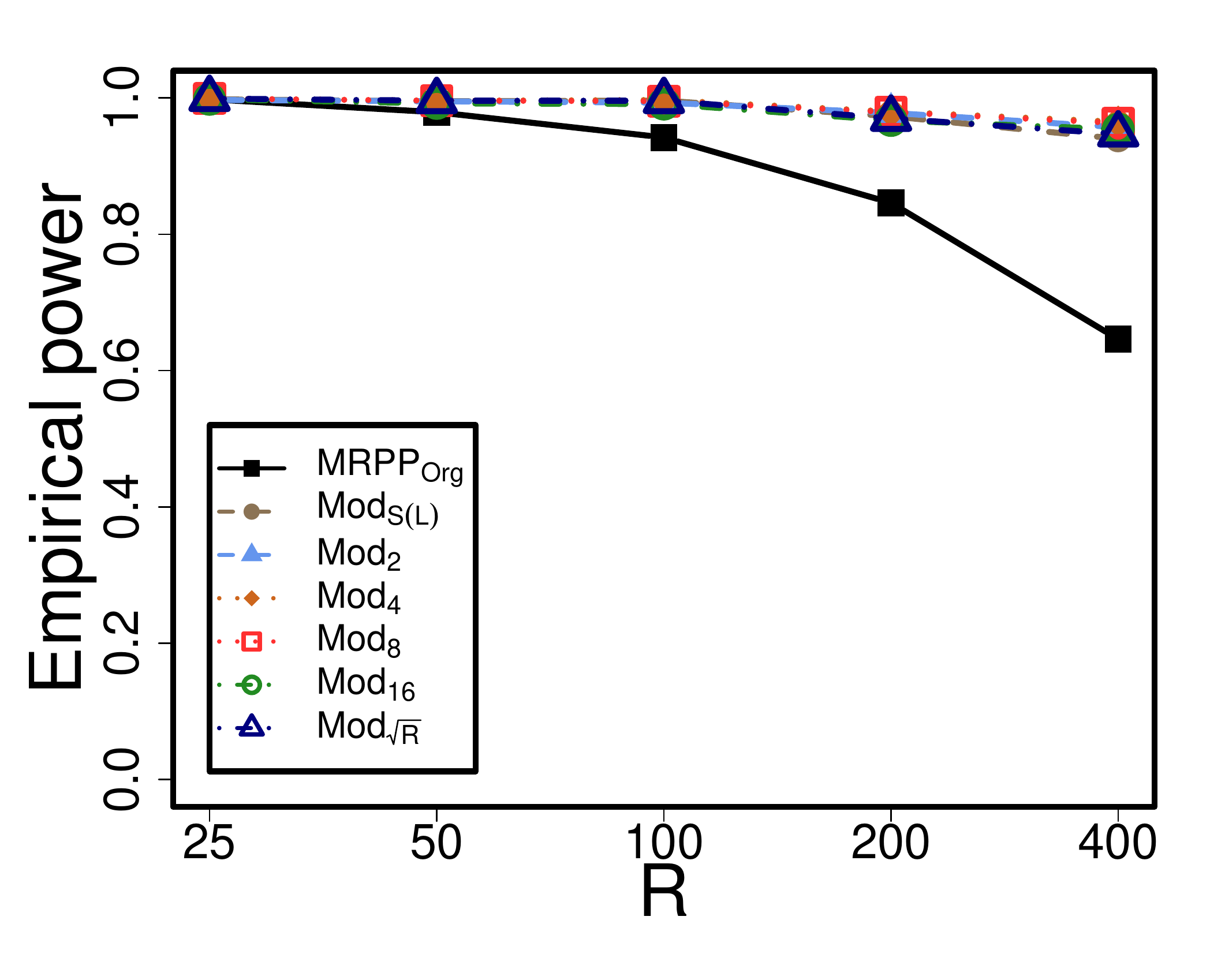}
		\label{fig:power_10_n1_40_n2_40}}
	\quad
	\subfigure[$n_1=40$, $n_2=80$]{%
		\includegraphics[width=0.45\linewidth]{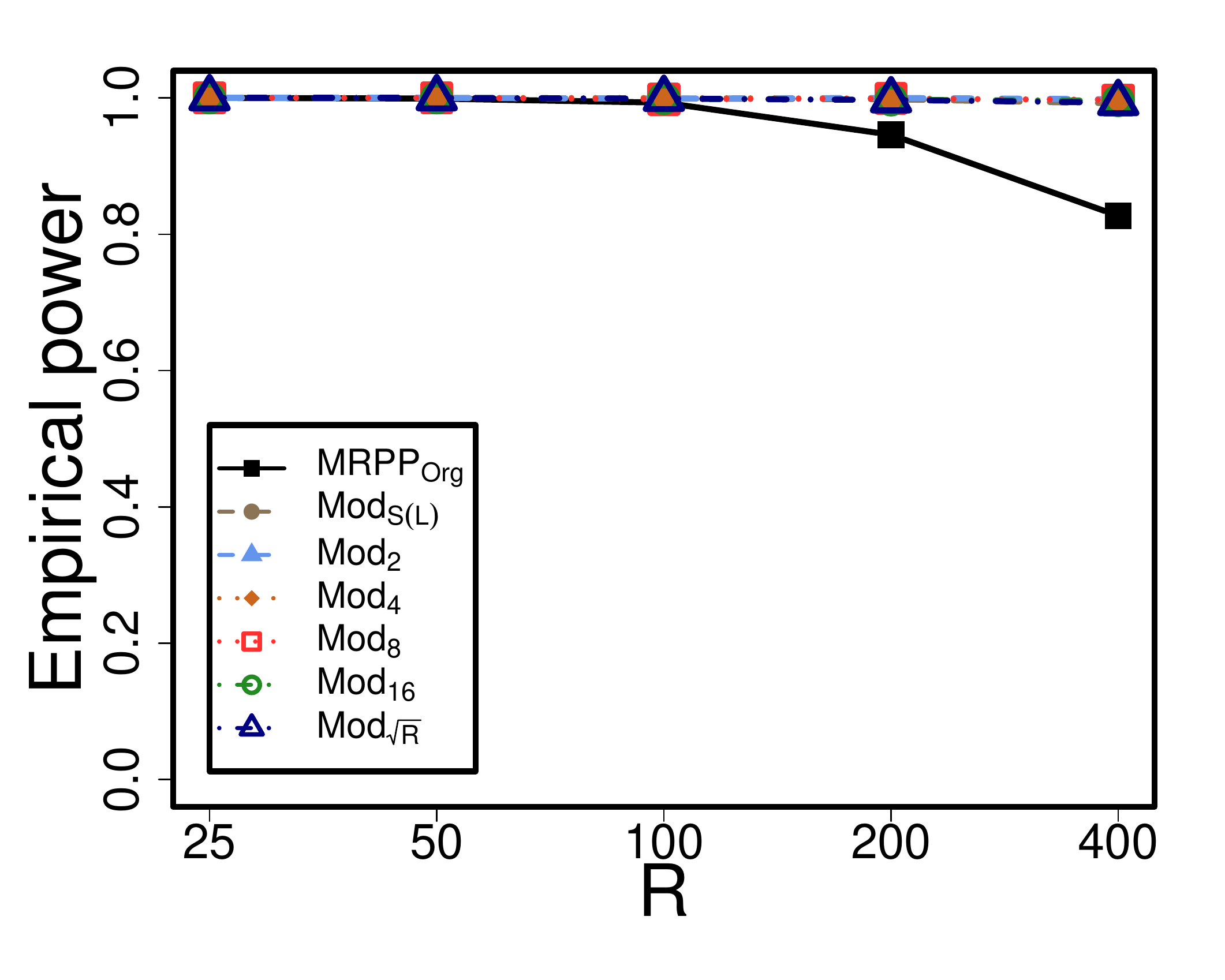}
		\label{fig:power_10_n1_40_n2_80}}
	\quad
	\subfigure[$n_1=80$, $n_2=80$]{%
		\includegraphics[width=0.45\linewidth]{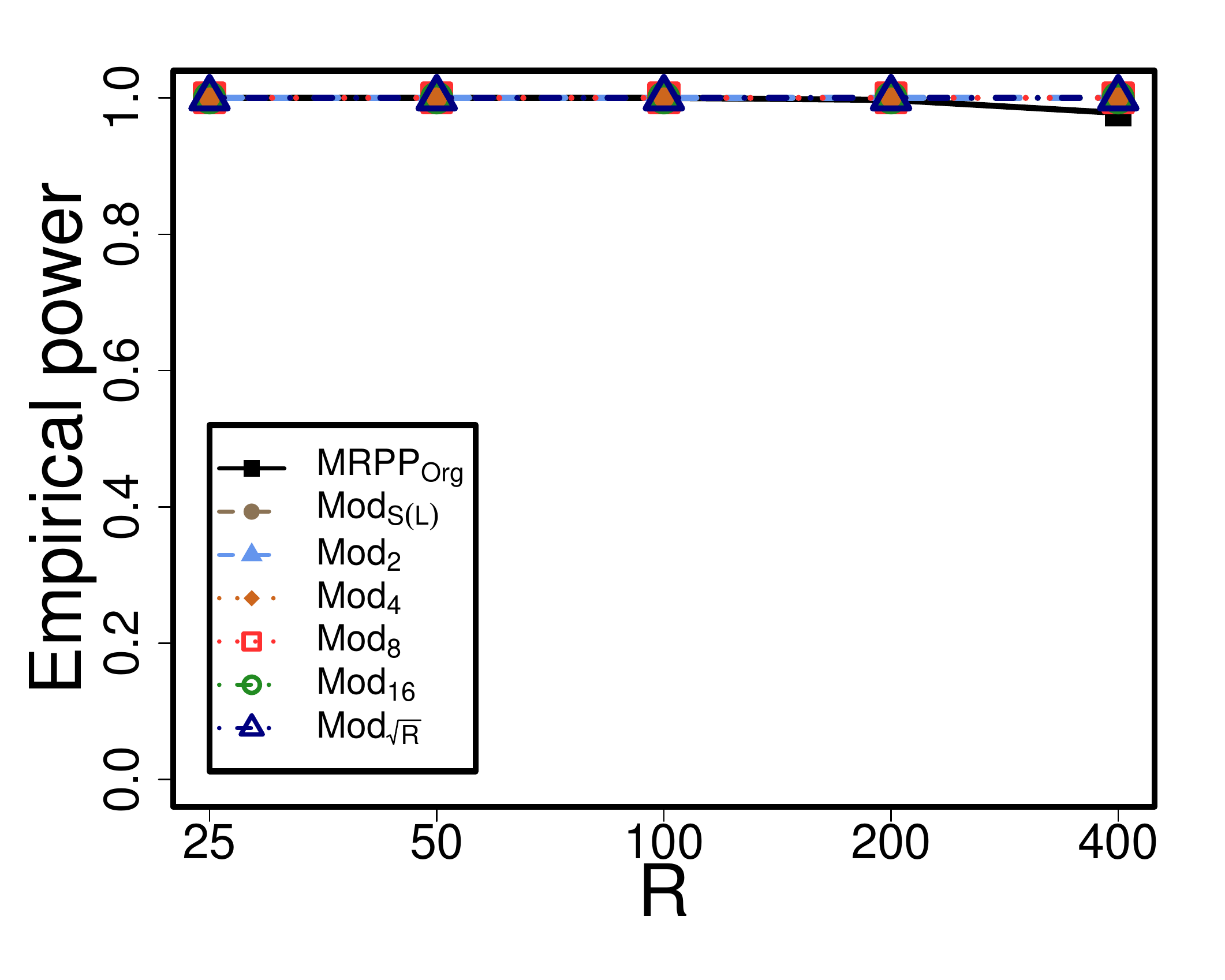}
		\label{fig:power_10_n1_80_n2_80}}
	\quad
	\caption{Empirical power of the original MRPP and modified MRPP with different choices of $R_0$ for $\nu=1$}
	\label{fig:modified_MRPP_power2}
\end{figure}

To investigate the power improvement of the modified MRPP test, for each combination of sample sizes $n_1$, $n_2$ and dimension $R$, we simulate $\{\bY_1, \ldots, \bY_{n_1}\}$ from $R$-dimensional multivariate normal distribution with mean vector $\mathbf{0}_R$ and covariance matrix $\bSigma=\left(0.5^{|i-j|}\right)_{1\leq i,j\leq R}$, while $\{\bY_{n_1+1},\ldots,\bY_N\}$ are drawn from the same distribution but with a location shift in the first four dimensions of its mean vector, that is, $\bmu=(\nu\mathbf{1}_4^T, \mathbf{0}_{R-4}^T)^T$ for the second group. The magnitude of location shift $\nu$ is chosen as $0.5$ and $1.0$. For the modified MRPP, the number of variables ($R_0$) used for testing is chosen in the same fashion as for our investigation of the test size. Figure \ref{fig:modified_MRPP_power1}--\ref{fig:modified_MRPP_power2} displays the empirical power of the original MRPP and the modified MRPP tests.

It is clear that power increases as $\nu$ increases for all testing approaches. In addition, as the sample sizes grow larger, all tests gain extra power as expected. The empirical power of the original MRPP decreases as the dimension $R$ increases, especially for the case when $\nu=1.0$. The empirical power of the modified MRPP also decreases when $R$ grows, but the decrease is much slower than that of the original MRPP. Because only a subset of important variables are used for testing the differences between two distributions based on our backward selection algorithm, the modified MRPP exhibits noticeable gains in power relative to the original MRPP for the largest $R$ settings. The empirical results suggests that the choice of $R_0$ does not have too much impact on the performance of the modified MRPP for the simulation scenarios we considered. 

\renewcommand{\baselinestretch}{1}

\renewcommand{\baselinestretch}{1}


\section{Discussion}

In this paper, we introduced importance measures based on MRPP and energy distance. The importance measures quantify the contribution of each variable in the difference between multivariate distributions. We developed a backward selection algorithm to address the variable selection problem for high-dimensional data. We examined the proposed backward selection approach by numerical studies and illustrated its applications in real data analysis. Furthermore, we modified the original MRPP using our proposed backward selection algorithm. Empirical evidence shows that the modified MRPP can not only preserve the nominal significance level, but also improve the power of the original MRPP by concentrating on the subset of most important variables when many variables are unimportant.

\section{Acknowledgment}

This material is based upon work supported by the National Science Foundation under Grant No. 1313224.

\end{document}